\documentclass[10pt,amsmath,amssymb,nofootinbib,twoside,twocolumn,superscriptaddress,floats,floatfix,aps,preprintnumbers]{revtex4-2}
\usepackage[british]{babel}
\usepackage{booktabs}
\usepackage{latexsym}
\usepackage{amssymb}
\usepackage{amsmath}
\usepackage{graphicx}
\usepackage{tabularx}
\usepackage{cancel}
\usepackage{hyperref}
\usepackage{colortbl}
\usepackage{aas_macros}
\usepackage{multirow}
\usepackage[capitalise]{cleveref}
\usepackage[utf8]{inputenc}
\usepackage[normalem]{ulem}
\usepackage[dvipsnames]{xcolor}

\begin{document}

\newcommand{\Salamanca}{\affiliation{Departamento de F\'isica Fundamental, Universidad de Salamanca, Plaza de la Merced, s/n, E-37008 Salamanca, Spain}}
\newcommand{\SalamancaIUFFyM}{\affiliation{Insituto Universitario de F\'isica Fundamental y Matem\'aticas (IUFFyM), Universidad de Salamanca, Plaza de la Merced, s/n, E-37008 Salamanca, Spain}}

\author{R.~Della~Monica}
\email{rdellamonica@tecnico.ulisboa.pt}
\affiliation{CENTRA, Departamento de Física, Instituto Superior Técnico – IST\\
Universidade de Lisboa – UL, Avenida Rovisco Pais 1, 1049-001 Lisboa, Portugal}
\Salamanca

\author{I.~De~Martino}
\email{ivan.demartino@usal.es}
\Salamanca
\SalamancaIUFFyM

\title{Pulsar timing in the Galactic Center}

\begin{abstract}
We propose a novel approach which implements the relativistic calculations of the photon travel time into a robust timing model for pulsars orbiting supermassive black holes. We demonstrate that timing models relying on the lowest-order (1PN) post-Newtonian approximation can produce significant discrepancies in strong-field configurations, affecting the predicted relativistic times of arrival at an Earth-based observatory. We also show how a misestimation of the pulsar parameters can lead to the appearance of phase-dependent residual, which hints at a tremendous constraining power of the binary and intrinsic parameters for timing observations of potential pulsars at the Galactic Center.
\end{abstract}

\maketitle

\section{Introduction}
\label{sec:intro}

The next significant advancement in experimental gravitation is expected with the discovery of pulsars orbiting a supermassive black hole (SMBH). Pulsars, whose intrinsic rotational period variation is of the order of one part in $10^{15}$ per pulse period \cite{Becker2018}, are a remarkable tool for probing gravitational fields \cite{Stairs2003, Lorimer2008, Will2014}. Pulsars in stellar binary systems allowed the detection of various general relativistic effects, including the decay of orbital periods due to the emission of {quadrupolar} gravitational waves \cite{Hulse1975, Taylor1994, Kramer2006a}. However, these systems probe a relatively weak gravitational field with compactness $GM/Rc^2\sim10^{-5}\div {10}^{-7}$ \cite{Zhang2017a}. Here $R$ represents the binary separation, and the component masses are on the order of solar masses. In contrast, a pulsar closely orbiting (\emph{i.e.} with orbital periods ranging from 1 to 100 years) a SMBH, such as Sagittarius A* (Sgr A*) at the Galactic Center \cite{DeLaurentis2023}, would explore an entirely different gravity regime where the source mass of the gravitational potential well is of order of $\sim 10^6M_\odot$. 

The significance of pulsars in the Galactic Center is highlighted by the increasing number of radio pulsar searches within the central few parsecs of the Milky Way \cite{Johnston1995, Johnston2006, Deneva2009, Deneva2010, Bates2011}. However, despite observational efforts, only five pulsars were discovered within 15 arcminutes of Sgr A* \cite{Johnston2006, Deneva2009}, along with a single radio magnetar positioned 2.4 arcseconds away from Sgr A* (equivalent to 0.1 pc in projection) \cite{Kennea2013, Mori2013, Rea2013}, and a millisecond pulsar within 1$^{\circ}$ from the Galactic Center \cite{Lower2024}. This apparent failure to detect pulsars at the Galactic Center is attributed to interstellar scattering processes. In fact, photons travel through the turbulent and ionized interstellar medium located between the source and the observer, which broadens the pulses \cite{Cordes2002}. Such a temporal broadening strongly depends on the observing frequency {($\propto\nu^{-\alpha}$ with $\alpha = 4$ for inhomogeneities in the scattering medium
following a Gaussian power spectrum and $\alpha = 4.4$ for a Kolmogorov spectrum \cite{Rickett1977})} decreasing the effectiveness of the periodicity search techniques even for long-period pulsars. Unlike dispersion, temporal broadening cannot be fully corrected in real time at the instrument level \cite{Eatough2013a}, and its impact can only be mitigated through multi-frequency observations and modelling techniques \cite{Shannon2017}. A potential solution would be to look at higher frequencies, but the characteristic power-law spectra of pulsars ($\propto \nu^{\alpha}$ with $\alpha < 0$ \cite{Wharton2012}) implies that higher frequencies correspond to a lower intrinsic flux. For such a reason,  high-frequency pulsar searches in the Galactic Center have not reported the detection of new pulsars even using Event Horizon Telescope (EHT) 2017 data \cite{Torne2021, Torne2023}. Nevertheless, the abundant population of young and massive stars orbiting Sgr A* suggests the potential presence of pulsars originating from Supernovae explosions in this population's massive end \cite{Figer2009} and, depending on the specific population model, there would be between 100 and 1000 pulsars within the central parsec of the Galactic Center, with orbital periods $<100$ years (including around 100 pulsars with orbital periods $<10$ years) and over 10,000-millisecond pulsars \cite{Pfahl2004, Zhang2014, Rajwade2017, Chennamangalam2014}. Therefore, detecting at least one pulsar orbiting Sgr A* has become a scientific goal for forthcoming observational facilities, including the Square Kilometre Array\footnote{\url{https://www.skao.int/en}} (SKA) \cite{Keane2015}, the Five-hundred-meter Aperture Spherical Telescope\footnote{\url{http://fast.bao.ac.cn}} (FAST) \cite{Nan2011}, the next-generation Very Large Array\footnote{\url{https://public.nrao.edu/ngvla/}} (ngVLA) \cite{Bower2018}, and the next generation Event Horizon Telescope\footnote{\url{https://www.ngeht.org}} (ngEHT) \cite{EventHorizonTelescopeCollaboration2022a}. 

On the theoretical side, the analysis of Times Of Arrival (TOAs) of pulses emitted by pulsars, usually known as pulsar timing analysis, might potentially improve all previous tests of General Relativity in the strong field regime \cite{Wex1999, Liu2012}. A benchmark for strong-field tests of gravity is the double pulsar PSR J0737–3039A/B, where pulsar timing has enabled the measurement of seven post-Keplerian parameters and higher-order relativistic effects, all in excellent agreement with general relativity \cite{Kramer2021, Hu2022}, making the double pulsar the current gold standard and motivating extensions to even stronger-field systems such as pulsars orbiting supermassive black holes. In this context, pulsar timing analyses in the Galactic Center will enhance the mass determination precision of Sgr A*, and could provide measurements of the spin magnitude and orientation of the SMBH \cite{Liu2012, Zhang2017a} with precisions, at 95\% of the confidence level, on the order of ${10}^{-3}\div {10}^{-2}$ and $10^{-1}-1$ degrees, respectively. Furthermore, these analyses could yield measurements of the SMBH quadrupole moment \cite{Wex1999, Liu2012, Psaltis2016}, allowing a direct test of the no-hair theorem at the Galactic Center \cite{Christian2015, Izmailov2019}. In \cite{DellaMonica2023d} a methodology for the numerical computation of fully relativistic propagation times for photons emitted in a generic spherically symmetric spacetime was introduced. This was applied to estimate how a change in the fundamental nature of the central object or in the underlying theory of gravity might leave an observational signature in the timing residuals. Interestingly, the results suggest that the possible constraints that future pulsar observations at the Galactic Center would enable on the extra parameters encoding deviations from the general relativistic scenario can surpass by orders of magnitude the constraints arising from other probes in the Galactic Center \cite{DeLaurentis2023}. This further demonstrates the paramount importance of pulsars orbiting SMBH as strong-field probes of gravity and motivates scientific and technical efforts required to successfully discover and time this kind of objects. Forthcoming telescopes will benefit from a substantial increase in the collection areas and, among them, SKA has been demonstrated to potentially be able to achieve an accuracy on the pulse TOA measurement of the order of $\sigma_\textrm{TOA} \simeq 100$~$\mu$s for normal pulsars \cite{Eatough2015} at frequencies above 15 GHz \cite{Liu2012} even when all possible limiting effects on the precision of TOA measurements of young pulsars in the Galactic Center environment are considered ({\it i.e.} the signal-to-noise ratio of the measured pulses, the intrinsic pulse phase jitter and the changes in pulse shape caused by interstellar scintillation). 

We aim to introduce a different approach for studying the TOAs of pulsars orbiting the SMBH in the Galactic Center which is based on the methodology to compute the photon propagation time in black hole spacetimes introduced in \cite{DellaMonica2023d}. First, we will assume that pulsars at the Galactic Center actually exist and will be successfully detected and timed within the accuracy goal of $100\,\mu$s per TOA. With these hypotheses in mind, we will first summarize in Section \ref{sec:framework} the theoretical framework in which we will move. Then, in Section  \ref{sec:pipelines} we will explain how, given a set of pulsar parameters, we can estimate the corresponding TOAs received by a distant observer using both the standard pulsar timing techniques based on first-order post-Newtonian (1PN) approximations and using our novel methodology. Here we show that 1PN formulas can deviate appreciably from fully relativistic results in the strong-field configurations considered. Finally, we discuss our results in Section \ref{sec:results} and draw our conclusions in Section \ref{sec:conclusions}.

\section{Theoretical Framework}
\label{sec:framework}

\subsection{Spherically symmetric spacetimes in General Realtivity}

In General Relativity, whenever one wants to describe the exterior gravitational field of static and spherically symmetric bodies, one has to search for spherically symmetric solutions to Einstein's field equations in vacuum. Looking for spherically symmetric solutions implies that one is looking for a metric tensor $g_{\mu\nu}$ whose components must satisfy certain symmetries, thus reducing the degree of freedom of the metric tensor and, hence, the number of free functions to solve for. In particular, the spacetime must be stationary and static and the metric tensor reduces to $g_{\mu\nu} = g_{tt}dt^2 + g_{ij}dx^idx^j$. Therefore, a convenient system of coordinates to employ is given by $(t,\,r_s,\,\theta,\,\phi)$, usually called the \emph{spherical (or Schwarzschild) coordinate system} where the components of the metric tensor depend on $(\theta,\,\phi)$ only through the solid angle which, in turn, implies that all off-diagonal terms in $g_{ij}$ must be zero. This also gives that $g_{\theta\theta} = r_s^2$ and $g_{\phi\phi} = r_s^2\sin^2\theta$. Finally, there are only two unknown functions of the aerial radius to be determined (corresponding to the $g_{tt}$ and $g_{rr}$ components) and the metric tensor may be recast as
\begin{equation}
    g_{\mu\nu} = \begin{pmatrix}
    A(r_s) & 0 & 0 & 0\\
    0 & B(r_s) & 0 & 0\\
    0 & 0 & r_s^2  & 0 \\
    0 &  0 &  0 & r_s^2\sin^2\theta\\
    \end{pmatrix}
\end{equation}
which corresponds to the following metric elements
\begin{equation}
    ds^2 = -A(r_s) dt^2 + B(r_s) dr_s^2 + r_s^2d\Omega^2\,.
    \label{eq:spherically_symmetric_space_time}
\end{equation}
Here $d\Omega^2 = d\theta^2+\sin^2\theta d\phi^2$, and the unknown functions are $A(r_s)$ and $B(r_s)$.  Let us remark that the line element in Eq. \eqref{eq:spherically_symmetric_space_time} only descends from symmetries imposed to a generic spacetime metric, without accounting for the field equations of General Relativity. For this reason, this line element is rather generic and can be used for any metric theory of gravity \cite{DellaMonica2023d}. In General Relativity, the solution of Einstein's field equations in vacuum returns the well-known Schwarzschild solution (here and in the rest of the paper we assume $G=c=1$)
\begin{equation}
    A(r_s) = B^{-1}(r_s) = 1-\frac{2M}{r_s}
    \label{eq:sch-spacetime}
\end{equation}
which is unique, obeys Birkhoff's theorem, and has only one free parameter, $M$, which corresponds to the mass of the central body. The post-Newtonian approximation of the spherically symmetric spacetime of General Relativity can be directly derived from the Schwarzschild solution in the weak-field, slow-motion limit, at any given order. The usual procedure to do this, however, is conventionally performed in a different system of coordinates, namely, the \emph{harmonic coordinates}. To facilitate comparison between our results and those from the usual post-Newtonian approach to pulsar timing, we will adopt this system of coordinates, too. The Schwarzschild solution in Eq. \eqref{eq:sch-spacetime} can be recast in harmonic coordinates by simply considering the coordinate transformation
\begin{equation}
    r := r_s + M
\end{equation}
which leads to
\begin{equation}
    ds^2 = -\left(\frac{r-M}{r+M}\right)dt^2 + \left(\frac{r+M}{r-M}\right)dr^2+(r+M)^2d\Omega^2.
\end{equation}
Since our main tool will be the geodesic motion in such spherically symmetric spacetime solutions, it is convenient to explicitly write down the geodesic equations for the metric element in Eq. \eqref{eq:spherically_symmetric_space_time}
\begin{align}
    \ddot{t} &= -\frac{2 M \dot{r}\dot{t}}{r^2-M^2}\label{eq:spherically_symmetric_geodesic_1}\\
    \ddot{r} &= \frac{M\dot{r}^{2}}{r^2-M^2} - \frac{M \dot{t}^{2} \left(r-M\right)}{(r+M)^3} + (r-M)(\dot{\phi}^{2}\sin^{2}\theta + \dot{\theta}^{2})\\
    \ddot{\theta} &= \frac{{M \dot{\phi}^{2} \sin{\left(2 \theta \right)}} + {r \dot{\phi}^{2} \sin{\left(2 \theta \right)}} - 4 \dot{r} \dot{\theta}}{2(r + M)}\\
    \ddot{\phi} &= - \frac{2\dot{\phi} \left((r+M)\dot{\theta}\cot\theta + \dot{r}\right)}{r+M}\label{eq:spherically_symmetric_geodesic_2}
\end{align}
where dots denote derivatives with respect to the affine parameter. These equations describe both massive particles' trajectories (time-like geodesics) and photon paths (null geodesics), the only difference between the two cases arising from the different normalization of the 4-velocity ($g_{\mu\nu}\dot{x}^\mu\dot{x}^\nu = -1$ for the time-like case and $g_{\mu\nu}\dot{x}^\mu\dot{x}^\nu = 0$ for the null case). Nonetheless, in our code we will treat the two different cases separately. Eqs. \eqref{eq:spherically_symmetric_geodesic_1}-\eqref{eq:spherically_symmetric_geodesic_2} will be integrated numerically as they appear for massive particle, using our code \texttt{PyGRO} \cite{PyGRO2025}. To compute photon paths, and thus to derive the photon propagation time in the spacetime described in Eq. \eqref{eq:spherically_symmetric_space_time}, on the other hand, we will resort to a numerical methodology which requires recasting the geodesic equations into a system of first-order differential equations, making use of the constants of motion. This methodology was introduced in \cite{DellaMonica2023d} and we briefly summarise it in the next section for completeness.

It is worth mentioning that analytical solutions to the geodesic equations for the Schwarzschild case, Eqs. \eqref{eq:spherically_symmetric_geodesic_1}-\eqref{eq:spherically_symmetric_geodesic_2}, are known for both massive and massless particles (see e.g. \cite{Chandrasekhar1998, Hackmann2019}). Such solutions rely on the specific expression taken by the metric coefficients, Eq. \eqref{eq:sch-spacetime}. However, since our goal is to develop a methodology that is valid generically for any metric belonging to the family of spherically-symmetric spacetimes as appearing in Eq. \eqref{eq:spherically_symmetric_space_time}, we need to resort to a numerical treatment for the geodesic equations \cite{DellaMonica2023d}. For the same reason, in the following, except where we will refer specifically to the Schwarzschild solution, we will leave $A(r)$ and $B(r)$ as free functions.

\subsection{Light propagation in a spherically symmetric spacetime}

\begin{figure*}
    \centering
    \includegraphics[width = 1.99\columnwidth]{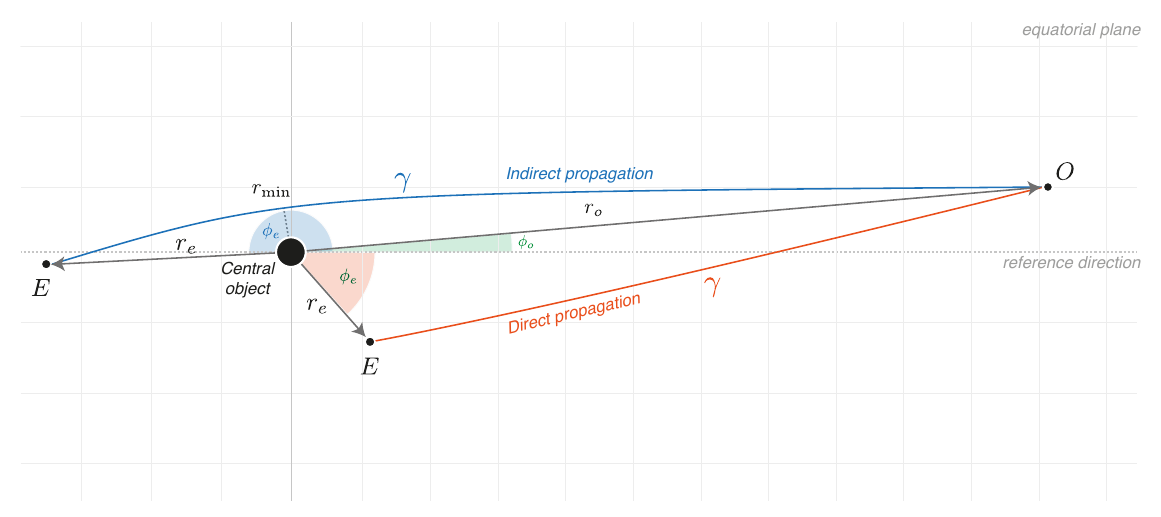}
    \caption{Illustration of the configuration for the emitter-observer problem. The emitter is located at a point $E$ that is identified by polar coordinates $(r_{\textrm{e}},\phi_{\textrm{e}})$, while the observer receiving the photon is located at point $O$ with coordinates $(r_{\textrm{o}} \phi_{\textrm{o}})$. Considering only primary photons received by the observer (\textit{i.e.} we do not consider photons that graze so close to the unstable photon orbit of the central object that their paths bend so  strongly, $\Delta \phi > 2\pi$, that they reach the observer after one or more complete turns around the central object) only two possible scenarios are possible: \textit{(green path)} the radial coordinate increases monotonically going from $r_{\textrm{e}}$ to $r_{\textrm{o}}$ \textit{propagating directly} ($r_{\textrm{e}}\to r_{\textrm{o}}$) from the emitter to the observer; \textit{(purple path)} the photon leaves the observer with a decreasing radial coordinate (\textit{i.e.} a negative radial velocity) then reaches a minimum distance $r_{\textrm{min}}$ from the central object after which it starts increasing again up to the observer position. We call the latter configuration \textit{indirect propagation} ($r_{\textrm{e}}\to r_{\textrm{min}} \to r_{\textrm{o}}$).}
    \label{fig:emitter_observer}
\end{figure*}

Relativistic effects on the propagation of photons are of paramount importance when dealing with pulsar timing. When null geodesics are integrated to describe the paths of photons, all these effects are automatically included. Nevertheless, to carry out the integration of the photon path, an appropriate initial condition search has to be carried out to identify the right null trajectory connecting the emitter and observer at any epoch or, in other words, to solve the \emph{emitter-observer} problem. In \cite{DellaMonica2023d}, a numerical methodology has been developed based on the generic asymptotically-flat spherically-symmetric spacetime in Eq. \eqref{eq:spherically_symmetric_space_time}. Since one is considering a spherical symmetric spacetime, one can restrict to the equatorial plane $\theta = \pi/2$ though this may not coincide with the plane on which the orbit of the emitting object lies. The propagation of the photon on such a plane is then determined by the equation for the polar trajectory of the photon,
\begin{equation}
    \frac{d\phi}{dr} = \frac{b}{r\sqrt{\Upsilon(r)}},
    \label{eq:dphi_dr}
\end{equation}
and the differential equation for the coordinate time, $t$, as a function of the radial coordinate, $r$,
\begin{equation}
    \frac{dt}{dr} = \frac{r}{A(r)\sqrt{\Upsilon(r)}}\,,
    \label{eq:dt_dr}
\end{equation}
where we have defined
\begin{equation}
    \Upsilon(r)=\left(\frac{r^2-b^2A(r)}{A(r)B(r)}\right)\,.
    \label{eq:upsilon}
\end{equation}
Here, we want to consider a pulsar at spatial coordinate $E(r_{\textrm{e}}, \phi_{\textrm{e}})$ which emits a photon $\gamma$ at coordinate time $t_{\textrm{e}}$, and an observer at spatial coordinate $O(r_{\textrm{o}}, \phi_{\textrm{o}})$ detecting the photon at coordinate time $t_{\textrm{o}}$. Therefore, Eq. \eqref{eq:dt_dr} is solved to compute the travel time, $\Delta t = t_{\textrm{o}}-t_{\textrm{e}}$, while Eq. \eqref{eq:dphi_dr} is used to solve for the impact parameter $b$ and, therefore, to solve the \emph{emitter-observer} problem. This latter step is done by solving the following equation
\begin{equation}
    \phi_{\textrm{o}}-\phi_{\textrm{e}} = 
    \int_{r_{\textrm{e}},\gamma}^{r_{\textrm{o}}}\frac{b}{r\sqrt{\Upsilon(r)}}dr.
    \label{eq:phi_integral}
\end{equation}
The integral in Eq. \eqref{eq:phi_integral} depends explicitly the path $\gamma$ taken by the photon going from $E$ to $O$. Since we will consider only primary photons received by the observer\footnote{We will not consider photons whose path bend so  strongly that they reach the observer after one or more complete turns around the central object} only two possible scenarios (illustrated in Figure \ref{fig:emitter_observer}) are possible:
\begin{description}
    \item[{\it Direct propagation}] the radial coordinate increases monotonically going from $r_{\textrm{e}}$ to $r_{\textrm{o}}$;
    \item[{\it Indirect propagation}] the radial coordinate decreases up to a certain $r_{\textrm{min}}$ and then increases monotonically up to $r_{\textrm{o}}$.
\end{description}
In the first case, the {explicit dependence on $\gamma$ appearing} in Eq. \eqref{eq:phi_integral} can result in the integral
\begin{equation}
    \phi_{\textrm{o}}-\phi_{\textrm{e}} = \int_{r_{\textrm{e}}}^{r_{\textrm{o}}}\frac{b}{r\sqrt{\Upsilon(r)}}dr,
    \label{eq:phi_integral_direct}
\end{equation}
while, in the other case, it would result in 
\begin{equation}
    \phi_{\textrm{o}}-\phi_{\textrm{e}} = \int_{r_{\textrm{e}}}^{r_{\textrm{min}}}\frac{b}{r\sqrt{\Upsilon(r)}} dr + \int_{r_{\textrm{min}}}^{r_{\textrm{o}}}\frac{b}{r\sqrt{\Upsilon(r)}}dr\,.\label{eq:phi_integral_indirect}
\end{equation}

Our numerical methodology, explained in detail in \cite{DellaMonica2023d}, allows to solve numerically for the unknown impact parameter $b$ appearing in Eq. \eqref{eq:phi_integral} and to determine the direct/indirect nature of the connecting photon, thus solving the emitter-observer problem.

We can now approach the problem of determining the travel time of the photon from $E$ to $O$. Since we consider line elements in Eq. \eqref{eq:spherically_symmetric_space_time} that are asymptotically flat, for a sufficiently far observer (as is the case for an Earth-based observer for a test particle orbiting Sgr A*), we can assume that the observer measures the coordinate time $t$ and thus the travel time is simply the integral of Eq. \eqref{eq:dt_dr} over the photon's path $\gamma$:
\begin{equation}
    \Delta t \equiv t_{\textrm{o}}-t_{\textrm{e}} = \int_{r_{\textrm{e}},\gamma}^{r_{\textrm{o}}} \frac{r}{A(r)\sqrt{\Upsilon(r)}}dr\,.
    \label{eq:relativistic_propagation_time}
\end{equation}
In the case of the Earth, one must add the relativistic corrections due to the presence of the Sun's gravitational field and the motion of Earth around it. Still, we can also safely assume a weak-field approximation for the Sun \cite{Damour1986}. Finally, considering the solution of the integral between two generic radial coordinates,
\begin{equation}
    T(r_1, r_2) =  \int_{r_1}^{r_2} \frac{r}{A(r)\sqrt{\Upsilon(r)}}dr\,,
    \label{eq:T_integral}
\end{equation}
we can express the direct and indirect propagation by
\begin{align}
    &\Delta t_{\textrm{direct}} = T(r_{\textrm{e}}, r_{\textrm{o}}),\\
    &\Delta t_{\textrm{indirect}} = T(r_{\textrm{e}}, r_{\textrm{min}})+T(r_{\textrm{min}}, r_{\textrm{o}}).
    \label{eq:delta_t_relativistic_pulsars}
\end{align}

As for the case of the emitter-observer problem, while an analytic solution in terms of Jacobi elliptic function exists for the case of the Schwarzschild spacetime, in the most general case one has to approach the problem numerically. 

\begin{figure}
    \centering
    \includegraphics[width = \columnwidth]{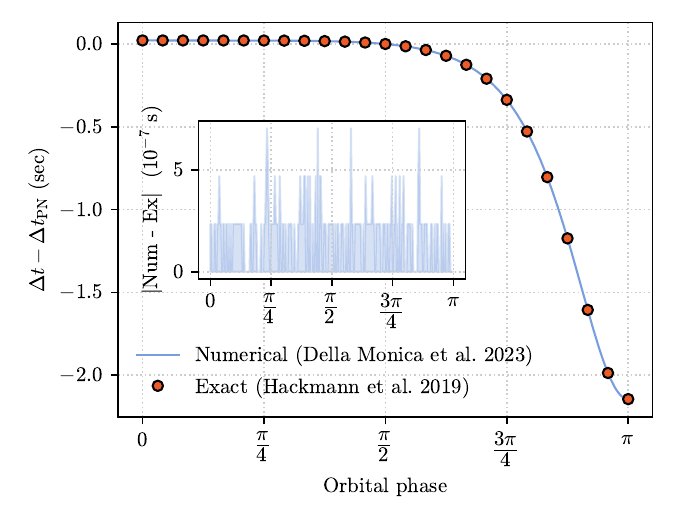}
    \caption{Residuals (in seconds) between the post-Newtonian approximation and the fully relativistic delay  (for both the exact formula as presented in \cite{Hackmann2019}, \emph{orange dots}, and our numerical methodology (blue solid line)) as a function of the orbital phase for a toy model circular orbit with $r = 100 M$ and an inclination of $i = 60^\circ$. In the configurations considered, the 1PN formula in Eq.~(27) does not fully reproduce the photon propagation time obtained from the fully relativistic treatment, with discrepancies reaching $\sim\,2$ s at superior conjunctions due to the increased time spent by the photon in the black hole gravitational field. The inset reports the residuals between the analytical estimation and the numerical one.}
    \label{fig:photon_propagation_exact_vs_numeric_PN}
\end{figure}

In \cite{Hackmann2019}, an analytical closed-form solution,  for a non-inclined circular orbit, has been found for a Schwarzschild spacetime in terms of Jacobi elliptic function. To show the effectiveness of our approach, we particularized the calculation of the time delay for the Schwarzschild spacetime and compared the numerical results with the exact solution given by \cite{Hackmann2019}
\begin{equation}
    \Delta t_{\textrm{ex}} = \frac{GM}{c^3}\left(T_{\textrm{ex}}(r_{\textrm{o}}; b)\pm T_{\textrm{ex}}(r_{\textrm{e}};b)\right).
    \label{eq:exact_formula}
\end{equation}
Here, the sign $\pm$ depends on whether the integration path is direct or indirect and the functions $T_{\textrm{ex}}(r;b)$ are, for a specific impact parameter, given by
\begin{align}
    T_{\textrm{ex}}(r;b) =& \frac{2}{\sqrt{r_4(r_3-r_1)}}(T_1+T_2+T_3+T_4) + T_\infty\,,
\end{align}
where we have defined the terms
\begin{align}
    T_1 &= \left(\frac{r_3^3}{r-2}+\frac{1}{2}(r_4-r_3)(r_3-r_1+4)\right)F(x,k),\\
    T_2 &= -\frac{1}{2}r_4(r_3-r_1)E(x,k),\\
    T_3 &= -2(r_4-r_3)\Pi\left(x,\frac{k^2}{c_1},k\right),\\
    T_4 &= -\frac{8(r_4-r_3)}{(r_4-2)(r_3-2)}\Pi(x,c_2,k)\,,
\end{align}
and $T_\infty$ encodes all the diverging (for $r\to\infty$) terms as follows
\begin{equation}
    T_\infty = \frac{b\sqrt{R(r)}}{r-r_3}+2\ln\left(\frac{\sqrt{r(r-r_1)}+\sqrt{(r-r_4)(r-r_3)}}{\sqrt{r(r-r_1)}-\sqrt{(r-r_4)(r-r_3)}}\right).
\end{equation}
In the previous relations, all quantities that appear are expressed in geometrized units, the functions $F$, $E$ and $\Pi$ are Jacobian elliptic integrals \cite{NISTDLMF}, $r_1$, $r_2$, $r_3$, $r_4$ are the roots of the function $R(r)$ which is derived from the function $\Upsilon(r)$ in Eq. \eqref{eq:upsilon}, upon factorization of the dependence on the impact parameter $b$, $x$ is an auxiliary variable and $k$, $c_1$ and $c_2$ are all constant terms built from the roots $r_1$, $r_2$, $r_3$, $r_4$. 

In Figure \ref{fig:photon_propagation_exact_vs_numeric_PN} we show \emph(i) the perfect agreement (within the adopted numerical tolerance of $10^{-6}$ s) between our numerical prediction and the analytical result for the Schwarzschild spacetime for a non-inclined circular orbit of radius 100$M$ and an observer at a distance of $10^9M$, which validate our approach and pipelines; and \emph(ii) the discrepancy that would arise by applying a 1PN approximation to compute the photon propagation time for the same orbital configuration. Let us remember that, at 1PN order, the photon travel time from the emitter's position to the observer's position can be written as a linear sum \cite{Damour1986,Edwards2006}
\begin{equation}
    \Delta t_{\textrm{PN}} = \Delta t_{\textrm{R}}+\Delta t_{\textrm{Sh}}+\Delta t_{\textrm{geo}}.
    \label{eq:post_newtonian}
\end{equation}
Here, $\Delta t_{ R}$ is the Rømer delay related to the photon propagation time across the emitting pulsar's orbit
\begin{equation}\label{eq:romer}
    \Delta t_{\textrm{R}}=\frac{|\vec{r}_{\textrm{o}}-\vec{r}_{\textrm{e}}|}{c}\,,
\end{equation}
and $\Delta t_{\textrm{Sh}}$ is the Shapiro time delay related to the time dilation of photons grazing the central object and is given by \cite{Will2014, Hackmann2019}:
\begin{equation}
    \Delta t_{\textrm{Sh}}=-\frac{2GM}{c^3}\ln\left(\frac{2|\vec{r}_{\textrm{o}}|}{|\vec{r}_{\textrm{e}}|+\vec{r}_{\textrm{e}}\cdot\hat{k}}\right)\,.
    \label{eq:shapiro}
\end{equation}
It is worth noticing that both the Rømer and the Shapiro delay are computed by assuming that the photons propagate on a straight line.
The last term, dubbed geometrical time delay, takes into account the curved photon path and is given by \cite{Will2014, Lai2005}
\begin{equation}
    \Delta t_\textrm{geo} = \frac{2GM}{c^3}\left(\frac{|\vec{r}_\pm-\vec{r}_s|}{R_E}\right)^2.
\end{equation}
where we have defined the gravitational-lensing Einstein radius
\begin{equation}
    R_E = \left(\frac{4GM}{c^2}(\vec{r}_\textrm{o}-\vec{r}_\textrm{e})\cdot\hat{k}\right)
\end{equation}
and the vector $\vec{r}_\textrm{s}=\vec{r}_\textrm{e}-(\vec{r}_\textrm{e}\cdot\hat{k}) \hat{k}$ is the sky-plane projection of the emitter position (\emph{i.e.} the apparent position that the pulsar would have in the observer reference frame without light bending), while
\begin{equation}
    \vec{r}_\pm=\frac{\vec{r}}{2}\left(1\pm\sqrt{1+\frac{4R^2_E}{|\vec{r}_\textrm{s}|}}\right)\,,
\end{equation}
is the sky-plane lensed position seen by the distant observer. 

Since the fully-relativistic photon propagation time, which can be derived analytically with equation Eq.~\eqref{eq:exact_formula} for the Schwarzschild spacetime or determined numerically with our approach for any spherically symmetric spacetime, captures strong field effects, it provides a better description of the photon propagation time with respect to the 1PN approximations in Eq. \eqref{eq:post_newtonian} when applied to a pulsar-SMBH system. When considering a circular orbit with a radius of about 100 gravitational radii, the two approaches differ by an amount $\sim\mathcal{O}(1s)$, as shown in Figure \ref{fig:photon_propagation_exact_vs_numeric_PN}. Moreover, in contrast to the standard PN timing models, which express the total time delay as a linear sum of contributions (Rømer, Shapiro, Einstein, etc.) for computational convenience, our fully relativistic framework does not introduce such a decomposition. Instead, all these effects are inherently included in the geodesic equations, so that the full non-linear interplay of general relativity is captured without approximation. This feature is particularly important in the strong-field regime, where non-linearities can become significant and may not be fully captured by a linear summation of separate terms.

Clearly, the specific expression for Eq. \eqref{eq:exact_formula} derived in  \cite{Hackmann2019} is only valid under the assumption of a Schwarzschild spacetime geometry and for a pulsar on a circular orbit. For a more general orbital model, or for different black hole geometries (even if spherical symmetry is preserved) such an expression should be modified accordingly and might even result in the impossibility of solving the corresponding integrals analytically. On the other hand, the approach developed in \cite{DellaMonica2023d} and summarised in this section is more general and allows the computation of the propagation times of photons for any spherically symmetric spacetime and any emitter-observer configuration capturing all relativistic effects.

\section{Pulsar timing techniques and pipelines}\label{sec:pipelines}

Using pulsar timing techniques, one can measure the TOAs of pulses emitted by a pulsar far away from the observer, monitor them on a timescale of years, and fit these TOAs to a model, namely a timing model. The latter is then used to link the measured TOA and the proper time of emission at the pulsar, allowing the computation of the pulse phase of emission accounting for the inherent variations in its period. Matching the predictions of the timing model with the observed TOAs, for pulsars in binary systems, one obtains a very precise estimation of the pulsar intrinsic parameters, \emph{e.g.} intrinsic period and spin-down rate, and of the orbital parameters, \emph{e.g.} masses and orbital period. 

Low-mass-ratio pulsar binary systems do not belong to a strong relativistic regime. Therefore, the evolution of such systems can be studied through a post-Newtonian treatment of both the orbital motion and photon propagation, which returns a timing model whose precision is below the experimental uncertainty on the TOAs. For this reason, all the pulsar timing codes available nowadays adopt this approach. Conversely, high-mass ratio binary pulsars with short orbital separations are in a strong gravitational field, and 1PN approximation of both the orbital motion and the photon propagation time might not provide a timing model that accurately reproduces the TOAs (an example of this is shown in Figure \ref{fig:photon_propagation_exact_vs_numeric_PN} {and will be analized in greater detail in Section~\ref{sec:post_newtonina_failure}}). If that is the case, the resulting residuals may bias parameter estimation when low-order approximations are used. Here, we propose a novel approach which implements the fully-relativistic calculations of the photon travel time, explained in the previous section, into a more robust timing model for pulsars orbiting SMBHs. We will perform a proof-of-concept analysis to investigate the inability of weak field 1PN approximation and the advantages of our methodology in the framework of parameter estimation for pulsars at the Galactic Centrer with SKA.

\subsection{Pulsar timing in the weak-field: current techniques and pipelines}
\label{sec:timing_tempo} 

First, let us summarize the standard pulsar timing techniques. Most of the known pulsar binary systems are characterised by a low mass ratio. For this reason, all the pulsar timing codes available nowadays adopt PN approximations to compute the timing model.  

\begin{figure*}
    \includegraphics[width=1.99\columnwidth]{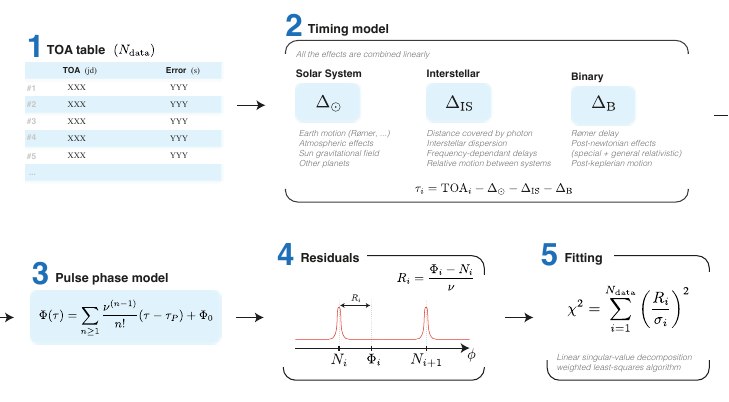}
    \caption{A schematic representation of the timing technique used in popular tools like \texttt{TEMPO2}. Observed TOAs are converted into barycentric proper emission times in the pulsar’s reference frame, $\tau$, using a timing model that accounts for various classical and relativistic delays. These emission times are then linked to the emitted pulse sequence by modeling the pulse phase as a Taylor series, incorporating the pulse period (or frequency), its temporal variation, and potentially higher derivatives. The timing residual, $R_i$, is the difference between the calculated pulse phase and the nearest integer $N_i$, expressed in time units by dividing by the frequency or multiplying by the period. A $\chi^2$ statistic is used to evaluate the alignment of the timing model with the observed TOAs. The best-fit timing model is determined by the parameters that minimize residuals and eliminate significant phase-dependent systematics.}
    \label{fig:tempo_timing}
\end{figure*}

The underlying mathematical framework and the accuracy in reproducing the TOAs are at the core of any code which aims to provide a timing model. Nevertheless, the rising sensitivity of radio observational facilities and the increasing diversity of pulsar populations discovered required improvements in the timing models accordingly.  For many years, a very popular choice for pulsar timing was \texttt{TEMPO}\footnote{\url{https://tempo.sourceforge.net}},
which provides 100 ns of timing accuracy for normal binary pulsar systems. Remarkably, it helped in numerous significant discoveries, such as the indirect detection of gravitational wave emission and other relativistic effects in pulsar binaries\cite{Taylor1979,Taylor1989}, the first detection of extrasolar planetary systems \cite{Wolszczan1992}, and the discovery of millisecond and sub-millisecond pulsars \cite{Kaspi1994,Edwards2001}. However, an updated version of \texttt{TEMPO}, namely \texttt{TEMPO2} \cite{Hobbs2006,Edwards2006}, was developed to reach the higher accuracy demanded by Pulsar Timing Array collaborations \cite{Agazie2023, EPTA2023, Reardon2023, Xu2023} to achieve the direct detection of a long-wavelength gravitational wave background. \texttt{TEMPO2} is capable of reaching $\sim1$ ns of timing accuracy for normal binary pulsar systems thanks to an improved characterization, with respect to \texttt{TEMPO}, of all the systematic effects that contribute to the resulting TOA, including tropospheric delays, pulse dispersion and updated estimates of secular orbital effects and Einstein and Shapiro delays. Recently, \texttt{PINT} \cite{Luo2021} based on the same theoretical principles of \texttt{TEMPO} and \texttt{TEMPO2}, was proposed to offer a modern coding environment that allows also for cross-checking the results obtained across the different implementations. The currently available timing codes model the binary motion at 1PN order using a post-Keplerian evolution of the orbital elements based on well established works in the literature \cite{Blandford1976,Damour1986,Taylor1989,Wex1999,Kopeikin1995,Kopeikin1996,Wex1998,Lorimer2004,Lorimer2004}. The photon propagation and hence the delays are computed at the same PN order. Recently, mainly motivated by the discovery of the double pulsar system PSR J0737–3039A/B, corrections of order 2PN and 2.5PN have been included on the orbital motion and terms up to 1.5PN have been considered for the photon delays \cite{Kramer2021, Hu2022}.

The whole timing procedure, which finally returns the physical parameters of the system and the estimation of the relativistic effects, is schematically depicted in Figure \ref{fig:tempo_timing}.  The first step is, of course, to measure the TOAs of pulses from a specific pulsar at a radio observatory. Such a measurement is done over an extended period (we will call $N_{\textrm{data}}$ the number of recorded TOAs for a specific pulsar), and must ensure that the observed TOAs accurately reflect arrival times in an inertial reference frame. Each measured arrival time must be transformed into the reference frame of the pulsar constructing the timing model. Then, if the residuals between the reconstructed pulses and the measured TOAs are not negligible and do not show a random structure, then the model is poorly determined and the estimation of the parameter is not robust. Such discrepancies would arise for unaccounted-for binary companions or parameters, irregularities in the pulsar's spin-down, or inaccurate estimation of astrometric or rotational parameters, among other effects\footnote{The full set of intrinsic, astrometric and binary parameters from which the timing model depends comprises more than 60 entries (albeit not all independent of each other) and is reported in Table A.1 of \cite{Edwards2006}.}.

In general, the timing model for a non-relativistic binary pulsar system is given by
\begin{equation}
    \tau_i = \textrm{TOA}_i - \Delta_\odot - \Delta_\textrm{IS} - \Delta_\textrm{B}\,,
    \label{eq:tempo_high_level_formula}
\end{equation}
where TOA$_i$ is the time of arrival at the observatory of the $i$-th pulse, $\tau_i$ is the proper time in the pulsar's inertial reference frame, $\Delta_\odot$ corresponds to the conversion to the Solar System barycenter frame, $\Delta_\textrm{IS}$ incorporates the transformation to the binary barycenter frame, and $\Delta_\textrm{B}$ accounts for the conversion to the pulsar frame. More specifically, $\Delta_\odot$ links the observed TOAs to the arrival time in the barycentric coordinate time at the Solar System barycenter, and accounts for vacuum propagation delays due to the Earth's orbital motion, its spin, precession and nutation, and any other delays due to the signal's passage through the Earth's atmosphere and the Solar system. This component links the observed TOAs to the arrival time in the barycentric coordinate time at the Solar System barycenter. $\Delta_\textrm{IS}$ incorporates delays attributed to the system's secular motion and due to the signal traversing the interstellar medium. Finally, $\Delta_\textrm{B}$ includes delays due to the binary orbital motion and the signal's passage through the gravitational field of the companion. 
The term $\Delta_\textrm{B}$ contains the relativistic contribution to the total photon propagation time produced in the binary pulsar system as a linear sum of post-Newtonian delays: the geometric Rømer delay in Eq. \eqref{eq:romer}, a pseudo-delay reflecting the aberration of the radio beam caused by binary motion, the Einstein delay that encompasses both gravitational redshift and special relativistic time dilation in the pulsar frame, and the Shapiro delay in Eq. \eqref{eq:shapiro}, which in this case takes the form
\begin{equation}
    \Delta_E = \gamma \sin u,
\end{equation}
where $\gamma$ is a theory dependent parameter (more details are reported in \cite{Damour1986, Edwards2006}) and $u$ is the pulsar's eccentric anomaly. When one considers an infinite mass ratio between the pulsar and the massive companion (as a pulsar around a SMBH), the term $\Delta_B$ reduces to the $\Delta t_\textrm{PN}$ introduced in Eq. \eqref{eq:post_newtonian}. {Finally, to describe the general relativistic motion of the binary, an orbital model based on the generalized post-Newtonian treatment of Damour and Deruelle \cite{Damour1986} is employed, which provides the orbital positions in the binary at 1PN order, also including secular derivatives for the orbital period, eccentricity and projected semimajor axis \cite{Blandford1976,Taylor1989,Wex1999,Kopeikin1995,Kopeikin1996,Wex1998,Lorimer2004,Lorimer2004}.}

Once the proper time is obtained by applying Eq. \eqref{eq:tempo_high_level_formula}, one can compute the corresponding pulse phase. In particular, if the pulsar period $P$ were perfectly constant (or, equivalently its angular frequency $\nu$), all the reconstructed proper times should be equally spaced and one would get integer steps from an initial phase $\Phi_0$. However, pulsars experience a rotational spin-down due to the magnetic-dipole braking \cite{Lorimer2004}. Therefore, the pulsar period is not perfectly constant (for all practical purposes one can consider only the first derivative, namely $\dot{\nu} = -\frac{\dot{P}}{P^2}$) implying a modification of the relation between the pulse phase $\Phi$ and the proper time of the pulses 
\begin{align}
    \Phi(\tau) &= \sum_{n\geq 1}\frac{\nu^{(n-1)}}{n!}(\tau-\tau_P)+\Phi_0 \nonumber\\
    & \approx \Phi_0 + \frac{\tau-\tau_P}{P} - \frac{\dot{P}}{2P^2}(\tau-\tau_P)^2,
    \label{eq:pulse_phase_model}
\end{align}
where a constant spin-down rate is considered. If one knew the true theoretical parameters used for the timing model, then the computed $\Phi_i-\Phi_0$ will be an integer number $N_i$ corresponding to the position of the given pulse in the pulse sequence (with the pulse at $\Phi_0$ corresponding to the 1st received). Nevertheless, the true parameters are not known \emph{a priori}, and data are affected by the measurement noise. Therefore, the deviations of these pulse phases from the nearest integer are finite, and are quantified  by the timing residual of the $i$-th pulse as
\begin{equation}
    R_i = \frac{\Phi_i-N_i}{\nu} = (\Phi_i-N_i)P
    \label{eq:pulsars_residuals}
\end{equation}
corresponding to the time since (or to) the nearest actual pulse emitted by the pulsar (see panel 4 in Figure \ref{fig:tempo_timing}). Thus, given a set of $N_\textrm{data}$ recorded TOAs for a specific pulsar, one can quantify the agreement of the timing model to the observational data by computing the $\chi$-squared
\begin{equation}
    \chi^2 = \sum_{i=1}^{N_\textrm{data}}\left(\frac{R_i}{\sigma_i}\right)^2,
    \label{eq:pulsars_chi2}
\end{equation}
where the $\sigma_i$ represent the observational uncertainties on the TOAs. Finally, one can apply the least-squares method iteratively to find the set of parameters that minimizes the $\chi$-squared. If those best-fit parameters do not produce any 
phase-dependent systematic in the residuals, the latter will only include measurement noise, and the timing model will return a robust estimation of both the pulsar's parameters and the relativistic effects experienced by the photon.

\subsection{Pulsar timing in strong-field regime}
\label{sec:timing_strong}

\begin{table}[!t]
    \footnotesize
    \setlength{\tabcolsep}{6.5pt}
    \renewcommand{\arraystretch}{1.7}
    \begin{tabular}{lccccc}
        \hline
        \textbf{Model} & $a$ (AU) & $a$ (mas) & $e$   & $T$  &   $\Gamma$ ($10^{-4}$)   \\   \hline
        {Toy 0} & 1025    & 125    & 0.88   & 16 yr  & 3\\
        {Toy 1} & 175.4    & 21.1      & 0.800   & 1.162 yr & 11 \\
        {Toy 2} & 80.0     & 9.75      & 0.800  & 126 days & 25 \\
        {Toy 3} & 43.8     & 5.28      & 0.800  & 52.9 days & 45 \\ \hline
    \end{tabular}
    \caption{The orbital parameters (semi-major axis in both physical units and angular dimension assuming a distance of $D = 8$ kpc for the Galactic Center, eccentricity and orbital period) of the toy models used for our analysis. The four orbits considered scan an increasingly strong gravitational regime, as quantified by the relativistic parameter $\Gamma\equiv r_g/r_p = GM/ac^2(1-e)$, corresponding to the ratio of the gravitational radius of the central object and the orbital separation of the pulsar at pericenter.}
    \label{tab:pulsars_toy_models}
\end{table}

\begin{figure*}[!ht]
    \includegraphics[width=1.99\columnwidth]{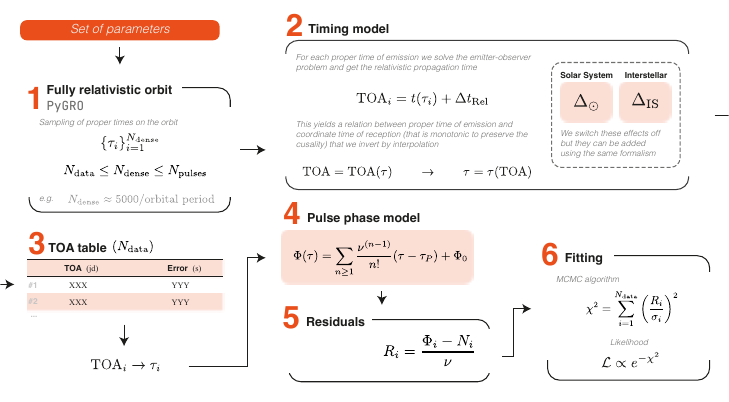}
    \caption{A schematic overview of the timing technique that we propose in this work. For a given set of intrinsic, astrometric, and orbital parameters of a pulsar orbiting a SMBH, we calculate a fully relativistic orbit by integrating the geodesic equation using \texttt{PyGRO} \cite{PyGRO2025}, Eqs. \eqref{eq:spherically_symmetric_geodesic_1}-\eqref{eq:spherically_symmetric_geodesic_2}. The orbit is sampled at $N_\textrm{dense}$ equally spaced points in proper time, $\{\tau_i\}_{i=1}^{N\textrm{dense}}$. At each point, the emitter-observer problem is solved to estimate the relativistic propagation time for a distant observer. This establishes a map between the observed TOA and the proper emission time, which is inverted using spline interpolation. The resulting function, $\tau(\textrm{TOA})$, is applied to the observed TOAs to determine the barycentric proper emission time in the pulsar's reference frame. Using standard timing techniques, we calculate the residuals, $R_i$, and the $\chi^2$ statistic, which defines a likelihood function, $\mathcal{L}$, employed for Bayesian parameter estimation.}
    \label{fig:pygro_timing}
\end{figure*}

In this work, we consider several putative pulsars on tight orbits around Sgr A*. The orbital elements of the toy models we include in our analysis are listed in Table \ref{tab:pulsars_toy_models}. This choice of the orbital elements is made with the idea of sampling the parameter space between the gravitational regime probed by the S2 star (the best observed in the S-stars cluster \cite{GravityCollaboration2020c, GravityCollaboration2024}), which coincides with our model Toy 0, and increasingly stronger gravitational regimes (as shown in the last column of Table \ref{tab:pulsars_toy_models}). In doing so, we also consider that, since the pulsar population in the Galactic Center is expected to be the remnant from the observed young stellar population \cite{Chen2023}, the orbits of the toy models we take into account must be outside a region where tidal disruption events for stars in the intense gravitational field of Sgr A* might occur. This corresponds to requiring that the orbital separation between the pulsar and Sgr A* at pericenter must be above the tidal radius $r_T \sim 60 r_g \approx 2.4 $ AU \cite{Rees1988,Waisberg2018,Chen2023,Generozov2025}. 

For such close orbits, a lowest-order (1PN) approximation may be insufficient to reproduce TOAs at the targeted precision, since higher-order relativistic contributions and nonlinear couplings are not included. Indeed, the entire weak-field methodology heavily relies on two assumptions:
\begin{enumerate}
    \item[\em (i)] the gravitational regime has to be sufficiently weak to consider that all relativistic effects can be combined linearly to compute the overall propagation time as in Eq. \eqref{eq:tempo_high_level_formula} \cite{Edwards2006};
    \item[\em (ii)] the approximation made by truncating at 1 PN order the description of the motion of both the binary system and photons leads to errors below the accuracy goal ($\sim 1$ ns) and, most importantly, much smaller than the pulsar intrinsic period $P$.
\end{enumerate}

The latter condition is needed to interpret the timing residuals correctly. {If the timing model is not accurate enough to satisfy both conditions, then the $i$-th pulse within the pulse sequence cannot be correctly identified and the whole procedure loses its predictive capability \cite{Edwards2006}}. For instance, if one considers double-pulsar binaries, these deviations are already large compared to the 1 ns accuracy goal \cite{Edwards2006}, {requiring the inclusion of higher-order PN corrections (up to 2.5PN for orbital dynamics and 1.5PN for photon propagation) in the usual timing codes \cite{Kramer2021}}. Consequently, in more extreme cases, these deviations can become huge. Indeed, we have already shown that in the mildly relativistic pulsar-SMBH binary considered in Figure \ref{fig:photon_propagation_exact_vs_numeric_PN}, the deviations between the 1PN formulation of the photon propagation time and the geodesic approach can be as large as 2~s. This prompts a fully geodesic description, which requires incorporating the techniques developed for solving the emitter-observer problem \cite{DellaMonica2023d} and changing the way to generate a phase-connected estimation of the timing residuals. The whole process is schematically summarised in Figure \ref{fig:pygro_timing}, and it is composed of the following steps:
\begin{itemize}
    \item[\em (a)] Once the orbit has been integrated numerically (for which we use our open source code \texttt{PyGRO} \cite{PyGRO2025}), we sample a given number ($N_\textrm{dense}$) of equally spaced (in terms of proper time $\tau$) points $\{\tau_i\}_{i=1}^{N_\textrm{dense}}$. These are a subset of the whole sequence of all pulses ($N_\textrm{pulses}$) emitted over the considered timespan. For instance, considering a pulsar with a pulse period $P = 2$ s and an orbital model like Toy 1 in Table \ref{tab:pulsars_toy_models} we would have $N_\textrm{pulses} = 1.84\times10^7$ pulses emitted over every orbital period. 
    \item[\em (b)] For each of the sampled points, we solve the emitter-observer problem and obtain an estimate of the relativistic propagation time, $\Delta t_\textrm{relativistic}$ using Eq. \eqref{eq:delta_t_relativistic_pulsars}.
    \item[\em (c)] In such a way, we can compute the time of arrival at the distant observer for all the photons emitted at proper times $\{\tau_i\}_{i=1}^{N_\textrm{dense}}$, and it is given by
    \begin{equation}
        \textrm{TOA}_i = t(\tau_i) + \Delta t_{\textrm{Relativistic}, i},
    \end{equation}
    where $t(\tau_i)$ is the coordinate time of emission and it is computed directly from the 0-th geodesics equation. Moreover, the term $\Delta t_{\textrm{Relativistic}, i}$ only takes into account the relativistic contribution to the photon travel time.
    \item[\em (d)] Using the samples derived in the previous point, we reconstruct the timing function $\textrm{TOA}(\tau)$ between the proper time of emission in the pulsar's reference frame and the TOAs in the reference frame of a distant observer. The function $\textrm{TOA}(\tau)$ contains contributions from both classic and relativistic delays, and it is monotonically increasing because the principle of causality in General Relativity is preserved. The 1PN approximation of this function is given in Eq. \eqref{eq:tempo_high_level_formula}. We reconstruct the function $\textrm{TOA}(\tau)$ at $N_\textrm{dense}$ points numerically and we invert it by interpolation, obtaining an interpolating function for the inverse timing function $\tau = \tau(\textrm{TOA})$. We adopt a {quintic spline interpolator}, consistent with the order of convergence of the Runge-Kutta algorithm used for the geodesic integration {(see Appendix~\ref{app:interpolation} for more details on the interpolation procedure)}.
    \item[\em (e)] Using the inverse timing model obtained in the previous point, one can translate TOAs in the observer reference frame into proper times in the pulsar reference frame. After that one can proceed as with \texttt{TEMPO2}, computing the pulse phase, considering the first-order evolution of the pulsar period in Eq. \eqref{eq:pulse_phase_model}, and the residuals $R_i$ for each pulse as given in Eq. \eqref{eq:pulsars_residuals}.
    \item[\em (f)] Finally, one can build a $\chi$-squared statistic and fit the timing model to the mock catalogue of TOAs using an MCMC algorithm (in our case, we adopt the affine invariant ensemble sampler implemented in the Python package \texttt{emcee} \cite{ForemanMackey2013}).
\end{itemize}
It is important to remark a few caveats of the timing procedure just described. The number of points, $N_\textrm{dense}$, at which we sample the function  $\textrm{TOA}(\tau)$ fulfils the following condition
\begin{equation*}
    N_\textrm{data} \leq N_\textrm{dense} \leq N_\textrm{pulses}.
\end{equation*}
Here $N_\textrm{data}$ represents the number of measurements in our TOA dataset, and $N_\textrm{pulses}$ is the total number of pulses emitted by the pulsar over the considered timespan. The most accurate reconstruction of the function $\textrm{TOA}(\tau)$ would be obtained if $N_\textrm{dense} = N_\textrm{pulses}$, \emph{i.e.} a $1:1$ correspondence between emitted pulses, integrated photon paths, and computed TOAs at the observer's location. On the other hand, setting $N_\textrm{dense} = N_\textrm{data}$ would guarantee the fastest (in terms of computational time) option but would provide a robust reconstruction of the $\textrm{TOA}(\tau)$ function only if $N_\textrm{dense}$ computed photon TOAs correspond exactly to the pulses, in the pulses sequence, that are measured at the observatory. However, this information is not known \emph{a priori} and one cannot know to what position in the pulse sequence a given measured TOA corresponds. {By employing a convergence check on our interpolation procedure (explained in detail in Appendix~\ref{app:interpolation})}, we we have found that, for the toy models we have considered in this work, $N_\textrm{dense}$ roughly 500 times $N_\textrm{data}$ provides a robust reconstruction of the photon travel time profile. {The same convergence check can be performed as a preliminary step for each new system studied applying our methodology, providing an important way of calibrating the parameter $N_\textrm{dense}$ of the timing model.}

Additionally, the contributions from the interstellar medium ($\Delta_\textrm{IS}$), the gravitational interaction of the photon with the Solar System gravitational field ($\Delta_\odot$), and the clock and atmospheric corrections routinely included in \texttt{TEMPO2} are not considered at this stage in our model. {These terms are all well understood, belong to the weak-field regime, and can be added linearly \emph{a posteriori} (as is standard in \texttt{TEMPO2}). We emphasize, however, that their omission here is a limitation of the present proof-of-concept study, and their future implementation will be necessary to increase the realism of the simulated TOAs.}

\section{Results}
\label{sec:results}

Let us now summarize the results of our analysis and highlight the resulting differences with standard timing techniques based on the 1PN approach. First, in Section \ref{sec:results_strong} we show the results of a qualitative analysis of the residuals arising from a misestimation of one parameter at a time in our timing model. Then, in Section \ref{sec:post_newtonina_failure} we show how our fully relativistic timing model improves over the 1PN approximation in the strong gravitational regime.

\subsection{Qualitative analysis of parameter estimation sensitivity}
\label{sec:results_strong}

\begin{figure*}[ht]
    \includegraphics[width=1.99\columnwidth]{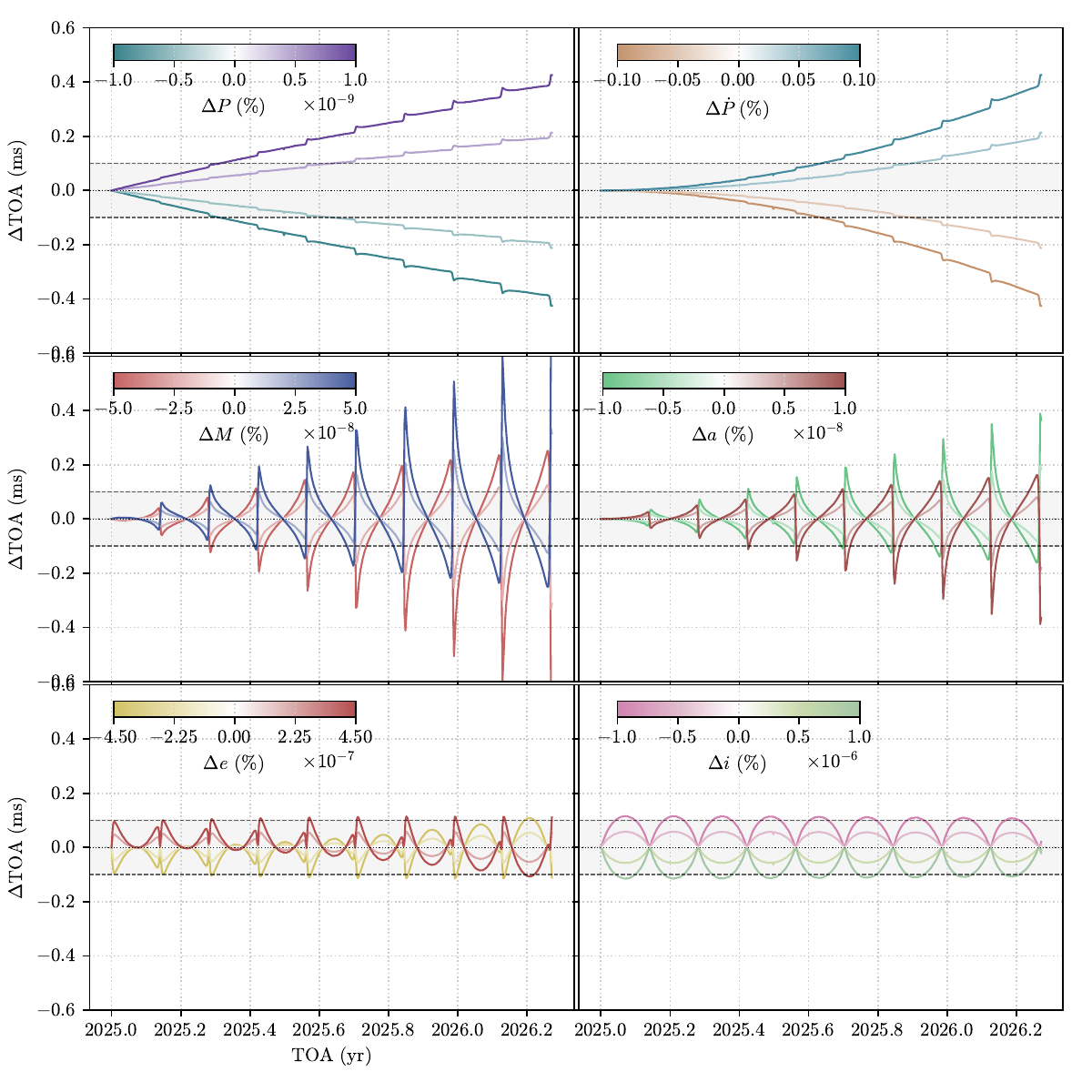}
    \caption{Timing residuals over 10 orbits for the Toy 3 model (see Table \ref{tab:pulsars_toy_models}), generated by slightly modifying each intrinsic or orbital parameter of the pulsar by the percentage indicated in the colorbar of each panel. Residuals are calculated by varying one parameter at a time while keeping all others fixed. The horizontal shaded regions represent the expected nominal timing accuracy for SKA observations in the Galactic Center, set at 100 $\mu$s.}
    \label{fig:residuals_toy3}
\end{figure*}

To appreciate how the shape and amplitude of the timing residuals change by over- (under-) estimating the orbital and intrinsic parameters of a pulsar, we perform a qualitative analysis of the TOA deviations. In Figure \ref{fig:residuals_toy3}, we show the timing residuals over 9 orbital periods for the Toy 3 model listed in table \ref{tab:pulsars_toy_models}.

In the top line, we considered the effect of a misestimation of the intrinsic pulsar period $P$ and its derivative $\dot{P}$ and show that both drastically alter the time of arrivals by linearly and quadratically drifting the residuals, respectively. More specifically, if we change the intrinsic pulsar period by only $\pm10^{-9}\%$  the timing residuals exceed the sensitivity threshold after only two orbital periods ($\sim 3$ months of observations). On the other hand, varying $\dot{P}$ of order $0.1\%$ makes the residuals exceed the sensitivity threshold after $\sim 4.5$  orbital periods ($\sim 7.5$ months of observations). 
We point out that changing intrinsic parameters, such as the pulsar period $P$ and its derivative $\dot{P}$, changes the shape and amplitude of the residuals but does not alter the orbit of the pulsar (nor the photon travel time). In fact, those changes in the shape and amplitude of the residuals do not depend on the particular system. For example, if we consider other toy models, the effect of varying $P$ and $\dot{P}$ will be the same, with the only difference being in the number of orbital periods required by their residuals to exceed the experimental detectability threshold.

Varying the mass $M$ of the gravitational source, panel (c) of Figure \ref{fig:residuals_toy3}, or of the orbital elements, panels (e) and (f) of Figure \ref{fig:residuals_toy3} for the semi-major axis $a$ and eccentricity $e$, respectively, affects the orbital dynamic itself of the pulsar. Therefore, secular effects will appear, accumulate and grow in amplitude on each orbit, leaving a detectable imprint on the shape and amplitude of the residuals. This is especially true for the mass of the central object $M$ and the semi-major axis of the pulsar orbits $a$, whose variations induce residuals with a linearly growing amplitude. {We remark that a constant perturbation to an orbital parameter (for instance $\delta M$) does not simply produce a fixed timing offset when the orbital phase evolution is considered. Since the orbital period depends on other orbital parameters (e.g. $T\propto M^{-1/2}$ in the Keplerian limit), a small parameter error alters the period by $\delta T/T$, so the predicted pulse phase drifts relative to the true phase as $\sim(\delta T/T)\,t$. Consequently, timing residuals obtained by comparing a model with misestimated orbital parameters to the true solution are expected to show secular growth. The secular trends in our numerical residuals are consistent with this effect; we therefore interpret them as physical phase-accumulation due to the incorrect orbital evolution.} When we vary the eccentricity $e$, on the other hand, the residuals show a constant amplitude at apocenter which scales proportionally to the semi-major axis (and does not increase over time), while the secular effect on the residuals only appears at apocenter (owing to the modified rate of orbital precession), whose amplitude exhibits a slower growth compared to the previous cases. Overall, this result clearly shows how the timing analysis of pulsars at the Galactic Center can not only provide incredibly tight constraints on the orbital elements of these objects, but also on the mass of the central SMBH which, with only a few months of observations, can surpass by orders of magnitudes the current precision from S-stars orbits.

Finally, in panel (g) of Figure \ref{fig:residuals_toy3}, we study the impact of changing the orbital inclination $i$ on the residuals. In this case, the amplitude is fixed because misestimating the angular orbital parameters in a spherically symmetric spacetime does not induce a secular variation of the orbits {and only results in phase-dependent residuals}. One must misestimate the inclinations by a factor $\sim10^{-6}\%$ to exceed the sensitivity threshold. 

By repeating this analysis for all the toy models considered in Table \ref{tab:pulsars_toy_models}, we obtain an estimate of the number of complete orbits (and hence of the observation timespan) required for the timing residuals to exceed the nominal timing sensitivity for SKA-like observations of pulsars at the Galactic Center. We summarize all these results in Table \ref{tab:residuals_all} for all the toy models reported in Table \ref{tab:pulsars_toy_models}. Specifically, for individual variations of one parameter at a time (by leaving all other parameters fixed) within given percentages, we report the number of periods and the corresponding time required for the residuals amplitude to surpass the threshold. The results are consistent across the four toy models considered, with pulsars in a stronger gravitational regime requiring a shorter observational window to qualitatively reach the same percentage sensitivity on the orbital parameters (with only the eccentricity, due to the aforementioned non-linear behaviour on the timing residuals, representing an exception to this pattern), while intrinsic parameters, like the pulsar period and its variation, whose estimation is independent on the strength of the gravitational regime, require the same amount of time across the different models.
Of course, the approach considered in this section, \emph{i.e.} considering variations of only one parameter at a time, does not take into account the possible degeneracies between the intrinsic and orbital pulsar parameters, which could not only spoil the precision of the parameter estimation, but also its accuracy (related, for example, to the introduction of biases in the parameter estimation).

\begin{table*}[!t]
    \footnotesize
    \renewcommand{\arraystretch}{2.5}
    \setlength{\tabcolsep}{8pt}
    {
    \begin{tabular}{lccccccccc}
        \hline
         \textbf{Parameter} & \textbf{Precision} & \multicolumn{2}{c}{\textbf{Toy 0} (S2-like)} & \multicolumn{2}{c}{\textbf{Toy 1}} & \multicolumn{2}{c}{\textbf{Toy 2}} & \multicolumn{2}{c}{\textbf{Toy 3}} \\ \hline
         &  & {Orbits} & {Time} & {Orbits} & {Time} & {Orbits} & {Time} & {Orbits} & {Time} \\ \hline
        \multicolumn{1}{c}{\multirow{2}{*}{$P$}} & $5\times10^{-10}\%$ & 0.03 & 6 months & 0.5 & 6 months & 1.9 & 239 days & 4 & 6 months \\
        \multicolumn{1}{c}{} & $1\times10^{-9}\%$ & 0.015 & 3 months & 0.25 & 3 months & 0.95 & 120 days & 2 & 3 months \\
         \multicolumn{1}{c}{} & $5\times10^{-2}\%$ & 0.05 & 10 months & 0.75 & 10 months & 2.6 & 328 days & 6 & 10 months \\
         \multicolumn{1}{c}{\multirow{-2}{*}{$\dot{P}$}} & $1\times10^{-1}\%$ & 0.04 & 7.5 months & 0.55 & 7.5 months & 1.85 & 234 days & 4.5 & 7.5 months \\
        \multicolumn{1}{c}{\multirow{2}{*}{$M$}} & $2.5\times10^{-8}\%$ & 0.73 & 11.6 years & 2 & 2.5 years & 2 & 250 days & 3 & 5 months \\ 
        \multicolumn{1}{c}{} & $5\times10^{-8}\%$ & 0.63 & 10 years & 1 & 1 year & 1 & 126 days & 2 & 3.5 months \\
         \multicolumn{1}{c}{} & $5\times10^{-9}\%$ & 0.71 & 11.3 years & 4 & 3.5 years & 4 & 500 days & 10 & 1.5 years \\
         \multicolumn{1}{c}{\multirow{-2}{*}{$a$}} & $1\times10^{-8}\%$ & 0.5 & 8.2 years & 3 & 2.5 years & 2 & 250 days & 3 & 5 months \\
        \multicolumn{1}{c}{\multirow{2}{*}{$e$}} & $2.5\times10^{-7}\%$ & 1 & 16 years & 1 & 1.2 years & 3.1 & 390 days & 20 & 2.4 years \\ 
        \multicolumn{1}{c}{} & $5\times10^{-7}\%$ & 1 & 16 years & 1 & 1.2 years & 1 & 126 days & 10 & 1.2 years \\
         \multicolumn{1}{c}{} &  & 0.5 & 8 years & 0.5 & 6 months & 0.5 & 63 days & 0.5 & 25 days \\
         \multicolumn{1}{c}{\multirow{-2}{*}{$i$}} & \multirow{-2}{*}{Case by case} & \multicolumn{2}{c}{\footnotesize ($5\times 10^{-8}\%$)} & \multicolumn{2}{c}{\footnotesize ($7\times 10^{-7}\%$)} &  \multicolumn{2}{c}{\footnotesize ($8\times 10^{-7}\%$)} & \multicolumn{2}{c}{\footnotesize ($1\times 10^{-6}\%$)} \\
        \hline
    \end{tabular}}
    \caption{The time (and corresponding number of complete orbits) required for the TOA residuals to surpass the nominal timing accuracy of 100 $\mu$s, expected for SKA observations of pulsars in the Galactic Center. For each toy model analyzed, the table lists the time needed to achieve the precision (in percentage) indicated in the second column for each of the pulsar's orbital and intrinsic parameters.}
    \label{tab:residuals_all}
\end{table*}

\subsection{Limitations of the 1PN timing formula in strong-field configurations}
\label{sec:post_newtonina_failure}

\begin{figure*}
    \includegraphics[width = \textwidth]{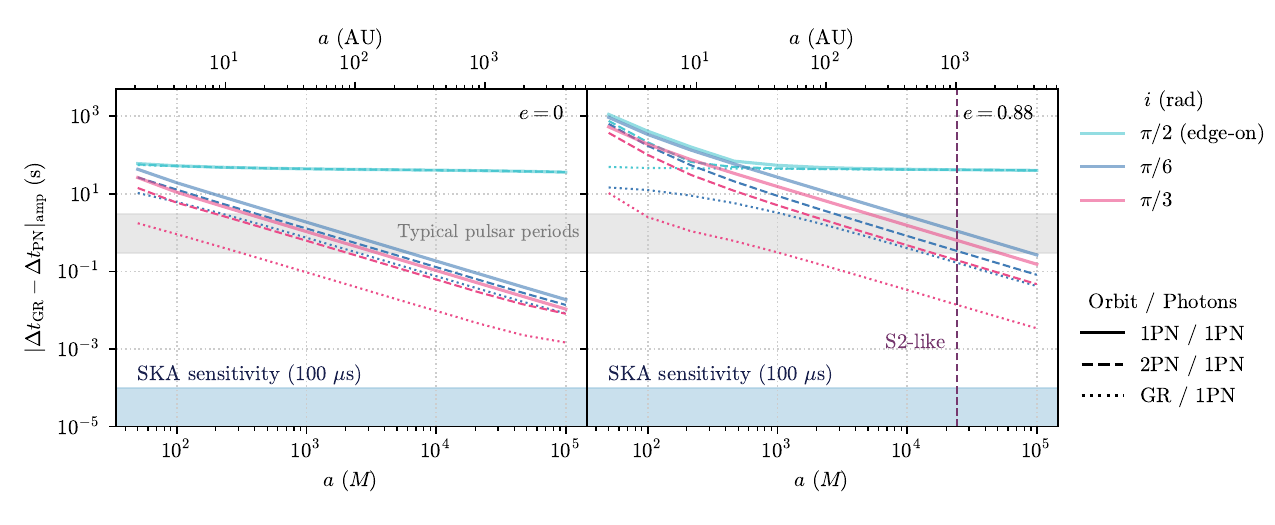}
    \caption{Amplitude of the residuals over one orbital period between the fully relativistic methodology and post-Newtonian timing formulas for pulsars around a SMBH. We consider the mass of the central object to be that of Sgr A* and two values of eccentricity, corresponding to circular orbits (\emph{left panel}) and to an S2-like orbit with $e= 0.88$ (\emph{right panel)} whose semi-major axis is highlighted with a dashed vertical line. We consider a range of semi-major axes spanning the interval $a\in[50,\,100000]M$ (corresponding to $[2,\,4200]$ AU for Sgr A*) and three values of the the inclination, as indicated in the legend. For each system we compute the residuals (of which the plots show the amplitude) between our fully relativistic model and several increasingly good levels of approximation: a full 1PN model (solid lines); one including a 2PN relativistic correction to the pericenter advance (dashed lines); a model based on a fully relativistic orbit and 1PN photon propagation estimates (dotted lines). The shaded regions corresponds to typical pulsar periods and the nominal SKA timing sensitivity.}
    \label{fig:pn_vs_pygro_timing}
\end{figure*}

The procedures introduced in the previous sections to compute TOAs for pulsars around a SMBH described by the Schwarzschild spacetime, also allows for an interesting comparison between results of the fully relativistic approach (described in Section \ref{sec:timing_strong}) and the ones obtained using formulas that rely on the post-Newtonian approximation, and are implemented in all current codes devoted to TOA analysis \cite{Edwards2006, Hobbs2006, Luo2021}, described in Section \ref{sec:timing_tempo}. Specifically, for a given set of orbital and intrinsic parameters, we compute a geodesic orbit for the pulsar covering one full orbital period, by numerical integration of Eqs. \eqref{eq:spherically_symmetric_geodesic_1}-\eqref{eq:spherically_symmetric_geodesic_2}. We sample emission proper times over the considered period and, at those positions on the orbit, we apply our fully relativistic methodology to compute the corresponding photon propagation time, $\Delta t_\textrm{GR}$, at the distant observer's location. At the same time, we consider a PN orbit as described in \cite{Damour1986} {with the same set of orbital parameters}. We then consider emission times corresponding to the same pulses in the pulse sequence as those used for the fully relativistic calculation, and, from the PN-predicted position of the pulsar at these instants of time, we compute the PN estimate of the photon propagation time, $\Delta t_\textrm{PN}$, using Eq. \eqref{eq:post_newtonian}. Since we are using harmonic coordinates for the geodesic integration, we can directly compare the two estimates of the propagation time, as the two are expressed in the same coordinate system. We do this by computing the amplitude of the residuals between the PN and the fully relativistic photon propagation time over the considered orbital period, defined as
\begin{align}
|\Delta t_\textrm{GR} - \Delta t_\textrm{PN}|_\textrm{amp} = &\max_T(\Delta t_\textrm{GR} - \Delta t_\textrm{PN})\nonumber\\
& - \min_T (\Delta t_\textrm{GR} - \Delta t_\textrm{PN}),
\label{eq:residuals_amplitude}
\end{align}
where the maximum and minimum are computed over the full orbital period $T$. Due to the way timing codes compute the pulse phase (and thus the residuals) from the coordinate TOA of a pulse (see Section \ref{sec:timing_tempo}), whenever the discrepancy that we compute surpasses the pulsar intrinsic period, it implies the failure of the timing procedure to correctly identify the emission pulse within the pulse sequence, over a period. In such cases, maintaining phase-connected residuals can become challenging without higher-order modeling or a fully relativistic treatment. We systematically analyze this residuals amplitude in Figure \ref{fig:pn_vs_pygro_timing}, where we report the quantity defined in Eq.~\eqref{eq:residuals_amplitude} over a wide range of orbital parameters. In particular, we include two cases: circular orbits (left panel) and a highly eccentric S2-like orbit with $e = 0.88$ (right panel). For both cases, residuals are calculated across a range of semi-major axes, $a \in [50,\,100000]M$ (or $[2,\,4200]$ AU when particularized for Sgr A*), and for three orbital inclinations. For each set of orbital parameters we compare our fully relativistic timing model with different PN timing models with increasingly better levels of description of the orbit. To fully disentangle the fraction of the residuals arising from the orbit and the photons, we consider: a full (orbit + photon propagation) 1PN model (solid lines) by directly implementing the Damour and Deruelle's model \cite{Damour1986}; a model incorporating 2PN relativistic correction to the pericenter advance \cite{Damour1988, Kramer2021} (dashed lines); a model based on our fully relativistic integration of the orbit and photon propagation computed through the usual 1PN formulas (dotted lines). Our results show that edge-on orbits exhibit the largest deviations due to pronounced strong lensing effects that are not correctly accounted for by 1PN photon propagation formulas. This is also confirmed by the fact that for this inclination, in th circular case all three orbital models result in overlapping amplitudes of the residuals, which are thus totally dominated by the inaccuracy of the 1PN photon propagation time when compared to the full relativistic geodesic approach. Inclined orbits show, as expected, amplitudes of the residuals that decrease with increasing distance of the pulsar from the central object (\emph{i.e.} towards a weaker gravitational regime) and with the better orbital models having smaller residuals overall. In broad strokes, less inclined and less eccentric orbits have smaller residual amplitudes than highly-inclined highly-eccentric orbits. We compare the amplitude of the residuals with typical pulsar periods from \cite{Manchester2005} and with the nominal timing accuracy of SKA, both reported as a shaded regions in the plot. Even for an orbit with same semi-major axis and eccentricity as the S2 star, the resulting discrepancy is of order $\sim \mathcal{O}(1\,\textrm{s})$, falling right in the region of usual pulsar periods and three orders of magnitude above the nominal SKA sensibility, and thus potentially able to spoil the ability to perform timing with current techniques. For smaller semi-major axes and more eccentric orbits, \emph{i.e.} in a stronger gravitational regime, this effect is even more drastic, easily exceeding typical pulsar periods by orders of magnitude for very tight orbits.

\section{Discussions and Conclusions}
\label{sec:conclusions}

The detection and precise timing of pulsars orbiting a SMBH, such as Sgr A* at the Galactic Center, represent an unparalleled opportunity to explore the strong-field regime of General Relativity and test alternative theories of gravity \cite{Liu2012, DellaMonica2023d}. Despite the challenges posed by interstellar scattering and observational limitations, the supposed existence of a substantial pulsar population near Sgr A* \cite{Figer2009, Pfahl2004, Zhang2014, Rajwade2017, Chennamangalam2014} strongly motivates continued efforts to detect and analyze these systems using advanced observational facilities like SKA \cite{Keane2015}, FAST \cite{Nan2011}, ngVLA \cite{Bower2018}, and ngEHT \cite{EventHorizonTelescopeCollaboration2022a}.

Our work builds upon this premise by introducing a novel approach to computing the photon propagation time in black hole spacetimes. This methodology, first introduced in \cite{DellaMonica2023d}, allows for more precise modeling of pulse TOAs at an Earth-based observatory and highlights the inadequacies, in these extreme environments, of traditional timing codes based on 1PN approximations. By comparing our relativistic approach with standard techniques, we demonstrate that our framework provides significantly improved accuracy, paving the way for more reliable measurements of the SMBH's mass (and possibly its spin and quadrupole moment), enabling robust tests of the black hole paradigm.

In Section \ref{sec:framework} we summarized the numerical methodology, first introduced in \cite{DellaMonica2023d} for the numerical computation of fully relativistic propagation times for photons emitted in a generic spherically symmetric spacetime, particularizing it for the Schwarzschild solution in harmonic coordinates, which are usually employed by timing codes based on the first-order post-Newtonian approximation (of which we have given an overview in Sectuon \ref{sec:timing_tempo}). We want to remark the fact that while the Schwarzschild solution is used as a benchmark for our numerical methodology, our pipeline, relying on \texttt{PyGRO} \cite{PyGRO2025}, benefits from full generality and can be applied to any spherically symmetric spacetime describing black holes (or mimickers) in GR or even in alternative theories of gravity.

In Sections \ref{sec:timing_strong} and \ref{sec:results_strong} we have extended our methodology not only to compute the general relativistic photon propagation time but also to develop a timing model to generate phase-connected residuals and perform fitting of TOAs data to a Schwarzschild model. Considering several toy model pulsars, we showed how this methodology can be used to quantitatively estimate the phase-dependent residuals arising from a misestimation of one of the pulsar's parameters. Such estimates, on the other hand, gave us a qualitative idea of the possible precision that would be achieved through an actual fitting analysis of our timing model on TOAs data. For example, we demonstrated a remarkable sensitivity of timing residuals to variations of the mass of central object as small as $\sim 10^{-8}\%$ of its value. This, even for an S2-like orbit, hints at the possibility of characterizing within only a few years of observations (or months, for more extreme orbital configurations) the gravitational field of the central SMBH of the Milky Way with a precision that is well beyond current estimates using orbital fitting of the S-Stars (current constrains on the mass of Sgr A* using these orbital data allow to determine its mass with a precision of $\sim 0.2\%$). The same also applies to the other orbital and intrinsic pulsar parameters.

On the other hand, in Section \ref{sec:post_newtonina_failure} we have shown how the fully relativistic estimates of the TOAs at a distant observer significantly differ from their corresponding estimates using both pulsar trajectories and photon propagation times computed with the same 1PN formulas used in popular pulsar timing codes \cite{Damour1986, Hobbs2006, Edwards2006}. This result supports the use of a fully relativistic methodology for computing TOAs in the strong field regime of a supermassive black hole and illustrates the limitations of the first order post Newtonian approximation in capturing nonlinear effects in the geodesic motion of both the pulsar and the emitted photons. Arguably, these issues might be alleviated extending current timing codes to include higher-order terms in the post-Newtonian expansion. A step forward in this direction has been done in recent years, mainly motivated by the discovery of the double pulsar system PSR J0737–3039A/B, and corrections of order 2PN and 2.5PN have been included on the orbital \cite{Kramer2021, Hu2022}. However, this quickly becomes impractical whenever one wants to test black hole models in theoretical frameworks beyond General Relativity. Our methodology, on the other hand, is fully general and can be applied to any spherically symmetric spacetime, providing fully-relativistic TOAs within any given numerical precision.

Our study highlights the great constraining power enabled by pulsar timing observations and the importance of properly accounting for all relativistic effects by adopting a geodesic timing model. The next major step to implement, in order to provide a more realistic description of the central object, is to extend the methodology that we have developed to a rotating black hole solution, like the Kerr black hole in General Relativity or its mimickers and extensions in modified theories of gravity.

\section*{Acknowledgements}
RDM and IDM acknowledge support from the grant PID2021-122938NB-I00 funded by MCIN/AEI/10.13039/501100011033 and by ``ERDF A way of making Europe''. RDM also acknowledges support from Consejeria de Educación de la Junta de Castilla y León, the European Social Fund + and financial support provided by FCT – Fundação para a Ciência e a Tecnologia, I.P., through the ERC-Portugal program Project ``GravNewFields''. RDM also thanks the Fundação para a Ciência e Tecnologia (FCT), Portugal, for the financial support to the Center for Astrophysics and Gravitation (CENTRA/IST/ULisboa) through grant No.~\href{https://doi.org/10.54499/UID/PRR/00099/2025}{UID/PRR/00099/2025} and grant No.~\href{https://doi.org/10.54499/UID/00099/2025}{UID/00099/2025}.. IDM also acknowledges support from the grant SA097P24 funded by Junta de Castilla y Le\'on and by "ERDF A way of making Europe". We finally thanks high-performance computing resources of the Castilla y Le\'on SupercomputingCenter (SCAYLE, \url{www.scayle.es} for providing supercomputing facilities.

\bibliography{biblio}

@article{Becker2018,
  author        = {{Becker}, Werner and {Kramer}, Michael and {Sesana}, Alberto},
  title         = {{Pulsar Timing and Its Application for Navigation and Gravitational Wave Detection}},
  journal       = {\ssr},
  year          = 2018,
  month         = feb,
  volume        = {214},
  number        = {1},
  eid           = {30},
  pages         = {30},
  doi           = {10.1007/s11214-017-0459-0},
  archiveprefix = {arXiv},
  eprint        = {1705.11022},
  primaryclass  = {astro-ph.IM},
  adsurl        = {https://ui.adsabs.harvard.edu/abs/2018SSRv..214...30B}
}

@article{Lorimer2008,
  author        = {{Lorimer}, Duncan R.},
  title         = {{Binary and Millisecond Pulsars}},
  journal       = {Living Reviews in Relativity},
  year          = 2008,
  month         = nov,
  volume        = {11},
  number        = {1},
  eid           = {8},
  pages         = {8},
  doi           = {10.12942/lrr-2008-8},
  archiveprefix = {arXiv},
  eprint        = {0811.0762},
  primaryclass  = {astro-ph},
  adsurl        = {https://ui.adsabs.harvard.edu/abs/2008LRR....11....8L}
}

@article{Hulse1975,
  author   = {{Hulse}, R.~A. and {Taylor}, J.~H.},
  title    = {{Discovery of a pulsar in a binary system.}},
  journal  = {\apjl},
  year     = 1975,
  month    = jan,
  volume   = {195},
  pages    = {L51-L53},
  doi      = {10.1086/181708},
  adsurl   = {https://ui.adsabs.harvard.edu/abs/1975ApJ...195L..51H}
}

@article{Taylor1994,
  author  = {{Taylor}, Joseph H., Jr.},
  title   = {{Binary pulsars and relativistic gravity}},
  journal = {Reviews of Modern Physics},
  year    = 1994,
  month   = jul,
  volume  = {66},
  number  = {3},
  pages   = {711-719},
  doi     = {10.1103/RevModPhys.66.711},
  adsurl  = {https://ui.adsabs.harvard.edu/abs/1994RvMP...66..711T}
}

@article{Kramer2006a,
  author        = {{Kramer}, M. and {Stairs}, I.~H. and {Manchester}, R.~N. and {McLaughlin}, M.~A. and {Lyne}, A.~G. and {Ferdman}, R.~D. and {Burgay}, M. and {Lorimer}, D.~R. and {Possenti}, A. and {D'Amico}, N. and {Sarkissian}, J.~M. and {Hobbs}, G.~B. and {Reynolds}, J.~E. and {Freire}, P.~C.~C. and {Camilo}, F.},
  title         = {{Tests of General Relativity from Timing the Double Pulsar}},
  journal       = {Science},
  year          = 2006,
  month         = oct,
  volume        = {314},
  number        = {5796},
  pages         = {97-102},
  doi           = {10.1126/science.1132305},
  archiveprefix = {arXiv},
  eprint        = {astro-ph/0609417},
  primaryclass  = {astro-ph},
  adsurl        = {https://ui.adsabs.harvard.edu/abs/2006Sci...314...97K}
}

@article{Stairs2003,
  author        = {{Stairs}, Ingrid H.},
  title         = {{Testing General Relativity with Pulsar Timing}},
  journal       = {Living Reviews in Relativity},
  year          = 2003,
  month         = {sep},
  volume        = {6},
  number        = {1},
  eid           = {5},
  pages         = {5},
  doi           = {10.12942/lrr-2003-5},
  archiveprefix = {arXiv},
  eprint        = {astro-ph/0307536},
  primaryclass  = {astro-ph},
  adsurl        = {https://ui.adsabs.harvard.edu/abs/Stairs2003}
}

@article{Will2014,
  author        = {{Will}, Clifford M.},
  title         = {{The Confrontation between General Relativity and Experiment}},
  journal       = {Living Reviews in Relativity},
  year          = 2014,
  month         = {dec},
  volume        = {17},
  number        = {1},
  eid           = {4},
  pages         = {4},
  doi           = {10.12942/lrr-2014-4},
  archiveprefix = {arXiv},
  eprint        = {1403.7377},
  primaryclass  = {gr-qc},
  adsurl        = {https://ui.adsabs.harvard.edu/abs/Will2014}
}

@article{Zhang2017a,
  author        = {{Zhang}, F. and {Saha}, P.},
  title         = {{Probing the Spinning of the Massive Black Hole in the Galactic Center via Pulsar Timing: A Full Relativistic Treatment}},
  journal       = {\apj},
  archiveprefix = {arXiv},
  eprint        = {1709.08341},
  year          = 2017,
  month         = {nov},
  volume        = 849,
  eid           = {33},
  pages         = {33},
  doi           = {10.3847/1538-4357/aa8f47},
  adsurl        = {http://adsabs.harvard.edu/abs/2017ApJ...849...33Z}
}

@article{Wex1999,
  author        = {{Wex}, N. and {Kopeikin}, S.~M.},
  title         = {{Frame Dragging and Other Precessional Effects in Black Hole Pulsar Binaries}},
  journal       = {\apj},
  year          = 1999,
  month         = mar,
  volume        = {514},
  number        = {1},
  pages         = {388-401},
  doi           = {10.1086/306933},
  archiveprefix = {arXiv},
  eprint        = {astro-ph/9811052},
  primaryclass  = {astro-ph},
  adsurl        = {https://ui.adsabs.harvard.edu/abs/1999ApJ...514..388W}
}

@article{Torne2021,
  author        = {{Torne}, P. and {Desvignes}, G. and {Eatough}, R.~P. and {Kramer}, M. and {Karuppusamy}, R. and {Liu}, K. and {Noutsos}, A. and {Wharton}, R. and {Kramer}, C. and {Navarro}, S. and {Paubert}, G. and {Sanchez}, S. and {Sanchez-Portal}, M. and {Schuster}, K.~F. and {Falcke}, H. and {Rezzolla}, L.},
  title         = {{Searching for pulsars in the Galactic centre at 3 and 2 mm}},
  journal       = {\aap},
  year          = 2021,
  month         = jun,
  volume        = {650},
  eid           = {A95},
  pages         = {A95},
  doi           = {10.1051/0004-6361/202140775},
  archiveprefix = {arXiv},
  eprint        = {2103.16581},
  primaryclass  = {astro-ph.HE},
  adsurl        = {https://ui.adsabs.harvard.edu/abs/2021A&A...650A..95T}
}

@article{Liu2012,
  author        = {{Liu}, K. and {Wex}, N. and {Kramer}, M. and {Cordes}, J.~M. and {Lazio}, T.~J.~W.},
  title         = {{Prospects for Probing the Spacetime of Sgr A* with Pulsars}},
  journal       = {\apj},
  year          = 2012,
  month         = mar,
  volume        = {747},
  number        = {1},
  eid           = {1},
  pages         = {1},
  doi           = {10.1088/0004-637X/747/1/1},
  archiveprefix = {arXiv},
  eprint        = {1112.2151},
  primaryclass  = {astro-ph.HE},
  adsurl        = {https://ui.adsabs.harvard.edu/abs/2012ApJ...747....1L}
}

@article{Psaltis2016,
  author        = {{Psaltis}, D. and {Wex}, N. and {Kramer}, M.},
  title         = {{A Quantitative Test of the No-hair Theorem with Sgr A* Using Stars, Pulsars, and the Event Horizon Telescope}},
  journal       = {\apj},
  archiveprefix = {arXiv},
  eprint        = {1510.00394},
  primaryclass  = {astro-ph.HE},
  year          = 2016,
  month         = {feb},
  volume        = 818,
  eid           = {121},
  pages         = {121},
  doi           = {10.3847/0004-637X/818/2/121},
  adsurl        = {http://cdsads.u-strasbg.fr/abs/2016ApJ...818..121P}
}

@article{Christian2015,
  author        = {{Christian}, Pierre and {Psaltis}, Dimitrios and {Loeb}, Abraham},
  title         = {{Shapiro Delays at the Quadrupole Order for Tests of the No-Hair Theorem Using Pulsars around Spinning Black Holes}},
  journal       = {arXiv e-prints},
  year          = 2015,
  month         = nov,
  eid           = {arXiv:1511.01901},
  pages         = {arXiv:1511.01901},
  doi           = {10.48550/arXiv.1511.01901},
  archiveprefix = {arXiv},
  eprint        = {1511.01901},
  primaryclass  = {gr-qc},
  adsurl        = {https://ui.adsabs.harvard.edu/abs/2015arXiv151101901C}
}

@article{Izmailov2019,
  author  = {{Izmailov}, Ramil N. and {Zhdanov}, Eduard R. and {Bhadra}, Arunava and {Nandi}, Kamal K.},
  title   = {{Relative time delay in a spinning black hole as a diagnostic for no-hair theorem}},
  journal = {European Physical Journal C},
  year    = 2019,
  month   = feb,
  volume  = {79},
  number  = {2},
  eid     = {105},
  pages   = {105},
  doi     = {10.1140/epjc/s10052-019-6618-6},
  adsurl  = {https://ui.adsabs.harvard.edu/abs/2019EPJC...79..105I}
}

@article{Johnston1995,
  author   = {{Johnston}, Simon and {Walker}, M.~A. and {van Kerkwijk}, M.~H. and {Lyne}, A.~G. and {D'Amico}, N.},
  title    = {{A 1500-MHz survey for pulsars near the Galactic Centre}},
  journal  = {\mnras},
  year     = 1995,
  month    = may,
  volume   = {274},
  number   = {2},
  pages    = {L43-L45},
  doi      = {10.1093/mnras/274.1.L43},
  adsurl   = {https://ui.adsabs.harvard.edu/abs/1995MNRAS.274L..43J}
}

@article{Johnston2006,
  author        = {{Johnston}, Simon and {Kramer}, M. and {Lorimer}, D.~R. and {Lyne}, A.~G. and {McLaughlin}, M. and {Klein}, B. and {Manchester}, R.~N.},
  title         = {{Discovery of two pulsars towards the Galactic Centre}},
  journal       = {\mnras},
  year          = 2006,
  month         = nov,
  volume        = {373},
  number        = {1},
  pages         = {L6-L10},
  doi           = {10.1111/j.1745-3933.2006.00232.x},
  archiveprefix = {arXiv},
  eprint        = {astro-ph/0606465},
  primaryclass  = {astro-ph},
  adsurl        = {https://ui.adsabs.harvard.edu/abs/2006MNRAS.373L...6J}
}

@article{Deneva2009,
  author        = {{Deneva}, J.~S. and {Cordes}, J.~M. and {Lazio}, T.~J.~W.},
  title         = {{Discovery of Three Pulsars from a Galactic Center Pulsar Population}},
  journal       = {\apjl},
  year          = 2009,
  month         = sep,
  volume        = {702},
  number        = {2},
  pages         = {L177-L181},
  doi           = {10.1088/0004-637X/702/2/L177},
  archiveprefix = {arXiv},
  eprint        = {0908.1331},
  primaryclass  = {astro-ph.SR},
  adsurl        = {https://ui.adsabs.harvard.edu/abs/2009ApJ...702L.177D}
}

@phdthesis{Deneva2010,
  author  = {{Deneva}, Julia Stefanova},
  title   = {{Elusive neutron star populations: Galactic center and intermittent pulsars}},
  school  = {Cornell University, New York},
  year    = 2010,
  month   = jan,
  adsurl  = {https://ui.adsabs.harvard.edu/abs/Deneva2010}
}

@article{Bates2011,
  author        = {{Bates}, S.~D. and {Johnston}, S. and {Lorimer}, D.~R. and {Kramer}, M. and {Possenti}, A. and {Burgay}, M. and {Stappers}, B. and {Keith}, M.~J. and {Lyne}, A. and {Bailes}, M. and {McLaughlin}, M.~A. and {O'Brien}, J.~T. and {Hobbs}, G.},
  title         = {{A 6.5-GHz multibeam pulsar survey}},
  journal       = {\mnras},
  year          = 2011,
  month         = mar,
  volume        = {411},
  number        = {3},
  pages         = {1575-1584},
  doi           = {10.1111/j.1365-2966.2010.17790.x},
  archiveprefix = {arXiv},
  eprint        = {1009.5873},
  primaryclass  = {astro-ph.SR},
  adsurl        = {https://ui.adsabs.harvard.edu/abs/2011MNRAS.411.1575B}
}

@article{Kennea2013,
  author        = {{Kennea}, J.~A. and {Burrows}, D.~N. and {Kouveliotou}, C. and {Palmer}, D.~M. and {G{\"o}{\u{g}}{\"u}{\c{s}}}, E. and {Kaneko}, Y. and {Evans}, P.~A. and {Degenaar}, N. and {Reynolds}, M.~T. and {Miller}, J.~M. and {Wijnands}, R. and {Mori}, K. and {Gehrels}, N.},
  title         = {{Swift Discovery of a New Soft Gamma Repeater, SGR J1745-29, near Sagittarius A*}},
  journal       = {\apjl},
  year          = 2013,
  month         = jun,
  volume        = {770},
  number        = {2},
  eid           = {L24},
  pages         = {L24},
  doi           = {10.1088/2041-8205/770/2/L24},
  archiveprefix = {arXiv},
  eprint        = {1305.2128},
  primaryclass  = {astro-ph.HE},
  adsurl        = {https://ui.adsabs.harvard.edu/abs/2013ApJ...770L..24K}
}

@article{Mori2013,
  author        = {{Mori}, Kaya and {Gotthelf}, Eric V. and {Zhang}, Shuo and {An}, Hongjun and {Baganoff}, Frederick K. and {Barri{\`e}re}, Nicolas M. and {Beloborodov}, Andrei M. and {Boggs}, Steven E. and {Christensen}, Finn E. and {Craig}, William W. and {Dufour}, Francois and {Grefenstette}, Brian W. and {Hailey}, Charles J. and {Harrison}, Fiona A. and {Hong}, Jaesub and {Kaspi}, Victoria M. and {Kennea}, Jamie A. and {Madsen}, Kristin K. and {Markwardt}, Craig B. and {Nynka}, Melania and {Stern}, Daniel and {Tomsick}, John A. and {Zhang}, William W.},
  title         = {{NuSTAR Discovery of a 3.76 s Transient Magnetar Near Sagittarius A*}},
  journal       = {\apjl},
  year          = 2013,
  month         = {jun},
  volume        = {770},
  number        = {2},
  eid           = {L23},
  pages         = {L23},
  doi           = {10.1088/2041-8205/770/2/L23},
  archiveprefix = {arXiv},
  eprint        = {1305.1945},
  primaryclass  = {astro-ph.HE},
  adsurl        = {https://ui.adsabs.harvard.edu/abs/Mori2013}
}

@article{Rea2013,
  author        = {{Rea}, N. and {Esposito}, P. and {Pons}, J.~A. and {Turolla}, R. and {Torres}, D.~F. and {Israel}, G.~L. and {Possenti}, A. and {Burgay}, M. and {Vigan{\`o}}, D. and {Papitto}, A. and {Perna}, R. and {Stella}, L. and {Ponti}, G. and {Baganoff}, F.~K. and {Haggard}, D. and {Camero-Arranz}, A. and {Zane}, S. and {Minter}, A. and {Mereghetti}, S. and {Tiengo}, A. and {Sch{\"o}del}, R. and {Feroci}, M. and {Mignani}, R. and {G{\"o}tz}, D.},
  title         = {{A Strongly Magnetized Pulsar within the Grasp of the Milky Way's Supermassive Black Hole}},
  journal       = {\apjl},
  year          = 2013,
  month         = {oct},
  volume        = {775},
  number        = {2},
  eid           = {L34},
  pages         = {L34},
  doi           = {10.1088/2041-8205/775/2/L34},
  archiveprefix = {arXiv},
  eprint        = {1307.6331},
  primaryclass  = {astro-ph.GA},
  adsurl        = {https://ui.adsabs.harvard.edu/abs/Rea2013}
}

@article{Lower2024,
  author        = {{Lower}, Marcus E. and {Dai}, Shi and {Johnston}, Simon},
  title         = {{The Snake's Beating Heart? A Millisecond Pulsar Binary in the Galactic Center Radio Filament G359.1$-$0.2}},
  journal       = {arXiv e-prints},
  year          = 2024,
  month         = apr,
  eid           = {arXiv:2404.09098},
  pages         = {arXiv:2404.09098},
  doi           = {10.48550/arXiv.2404.09098},
  archiveprefix = {arXiv},
  eprint        = {2404.09098},
  primaryclass  = {astro-ph.HE},
  adsurl        = {https://ui.adsabs.harvard.edu/abs/2024arXiv240409098L}
}

@article{Cordes2002,
  author        = {{Cordes}, J.~M. and {Lazio}, T.~J.~W.},
  title         = {{NE2001.I. A New Model for the Galactic Distribution of Free Electrons and its Fluctuations}},
  journal       = {arXiv e-prints},
  year          = 2002,
  month         = jul,
  eid           = {astro-ph/0207156},
  pages         = {astro-ph/0207156},
  doi           = {10.48550/arXiv.astro-ph/0207156},
  archiveprefix = {arXiv},
  eprint        = {astro-ph/0207156},
  primaryclass  = {astro-ph},
  adsurl        = {https://ui.adsabs.harvard.edu/abs/2002astro.ph..7156C}
}

@article{Eatough2013a,
  author        = {Eatough, R. P. and Kramer, M. and Klein, B. and Karuppusamy, R. and Champion, D. J. and Freire, P. C. C. and Wex, N. and Liu, K.},
  editor        = {van Leeuwen, Joeri},
  title         = {{Can we see pulsars around Sgr $A^⋆$? The latest searches with the Effelsberg telescope.}},
  eprint        = {1210.3770},
  archiveprefix = {arXiv},
  primaryclass  = {astro-ph.GA},
  doi           = {10.1017/S1743921312024209},
  journal       = {IAU Symp.},
  volume        = {291},
  pages         = {382--384},
  year          = {2013}
}

@article{Wharton2012,
  author        = {{Wharton}, R.~S. and {Chatterjee}, S. and {Cordes}, J.~M. and {Deneva}, J.~S. and {Lazio}, T.~J.~W.},
  title         = {{Multiwavelength Constraints on Pulsar Populations in the Galactic Center}},
  journal       = {\apj},
  year          = 2012,
  month         = jul,
  volume        = {753},
  number        = {2},
  eid           = {108},
  pages         = {108},
  doi           = {10.1088/0004-637X/753/2/108},
  archiveprefix = {arXiv},
  eprint        = {1111.4216},
  primaryclass  = {astro-ph.HE},
  adsurl        = {https://ui.adsabs.harvard.edu/abs/2012ApJ...753..108W}
}

@article{Torne2023,
  author        = {{Torne}, Pablo and {Liu}, Kuo and {Eatough} and other},
  title         = {{A Search for Pulsars around Sgr A* in the First Event Horizon Telescope Data Set}},
  journal       = {\apj},
  year          = 2023,
  month         = dec,
  volume        = {959},
  number        = {1},
  eid           = {14},
  pages         = {14},
  doi           = {10.3847/1538-4357/acf4f2},
  archiveprefix = {arXiv},
  eprint        = {2308.15381},
  primaryclass  = {astro-ph.HE},
  adsurl        = {https://ui.adsabs.harvard.edu/abs/2023ApJ...959...14T}
}

@article{Pfahl2004,
  author        = {{Pfahl}, Eric and {Loeb}, Abraham},
  title         = {{Probing the Spacetime around Sagittarius A* with Radio Pulsars}},
  journal       = {\apj},
  year          = 2004,
  month         = nov,
  volume        = {615},
  number        = {1},
  pages         = {253-258},
  doi           = {10.1086/423975},
  archiveprefix = {arXiv},
  eprint        = {astro-ph/0309744},
  primaryclass  = {astro-ph},
  adsurl        = {https://ui.adsabs.harvard.edu/abs/2004ApJ...615..253P}
}

@inproceedings{Eatough2015,
  author        = {{Eatough}, R. and {Lazio}, T.~J.~W. and {Casanellas}, J. and {Chatterjee}, S. and {Cordes}, J.~M. and {Demorest}, P.~B. and {Kramer}, M. and {Lee}, K.~J. and {Liu}, K. and {Ransom}, S.~M. and {Wex}, N.},
  title         = {{Observing Radio Pulsars in the Galactic Centre with the Square Kilometre Array}},
  booktitle     = {Advancing Astrophysics with the Square Kilometre Array (AASKA14)},
  year          = 2015,
  month         = apr,
  eid           = {45},
  pages         = {45},
  archiveprefix = {arXiv},
  eprint        = {1501.00281},
  primaryclass  = {astro-ph.IM},
  adsurl        = {https://ui.adsabs.harvard.edu/abs/2015aska.confE..45E}
}

@article{EventHorizonTelescopeCollaboration2022a,
  author   = {{Akiyama}, Kazunori and {Alberdi}, Antxon and {Alef}, Walter and {Algaba}, Juan Carlos and {Anantua}, Richard and {Asada}, Keiichi and {Azulay}, Rebecca and {Bach}, Uwe and {Baczko}, Anne-Kathrin and {Ball}, David and et al.},
  title    = {{First Sagittarius A* Event Horizon Telescope Results. I. The Shadow of the Supermassive Black Hole in the Center of the Milky Way}},
  journal  = {\apjl},
  year     = 2022,
  month    = {may},
  volume   = {930},
  number   = {2},
  eid      = {L12},
  pages    = {L12},
  doi      = {10.3847/2041-8213/ac6674},
  adsurl   = {https://ui.adsabs.harvard.edu/abs/Akiyama2022}
}

@article{Rajwade2017,
  author        = {{Rajwade}, K.~M. and {Lorimer}, D.~R. and {Anderson}, L.~D.},
  title         = {{Detecting pulsars in the Galactic Centre}},
  journal       = {\mnras},
  year          = 2017,
  month         = oct,
  volume        = {471},
  number        = {1},
  pages         = {730-739},
  doi           = {10.1093/mnras/stx1661},
  archiveprefix = {arXiv},
  eprint        = {1611.06977},
  primaryclass  = {astro-ph.HE},
  adsurl        = {https://ui.adsabs.harvard.edu/abs/2017MNRAS.471..730R}
}

@article{Chennamangalam2014,
  author        = {{Chennamangalam}, J. and {Lorimer}, D.~R.},
  title         = {{The Galactic Centre pulsar population.}},
  journal       = {\mnras},
  year          = 2014,
  month         = may,
  volume        = {440},
  pages         = {L86-L90},
  doi           = {10.1093/mnrasl/slu025},
  archiveprefix = {arXiv},
  eprint        = {1311.4846},
  primaryclass  = {astro-ph.HE},
  adsurl        = {https://ui.adsabs.harvard.edu/abs/2014MNRAS.440L..86C}
}

@inproceedings{Keane2015,
  author        = {{Keane}, E. and {Bhattacharyya}, B. and {Kramer}, M. and {Stappers}, B. and {Keane}, E.~F. and {Bhattacharyya}, B. and {Kramer}, M. and {Stappers}, B.~W. and {Bates}, S.~D. and {Burgay}, M. and {Chatterjee}, S. and {Champion}, D.~J. and {Eatough}, R.~P. and {Hessels}, J.~W.~T. and {Janssen}, G. and {Lee}, K.~J. and {van Leeuwen}, J. and {Margueron}, J. and {Oertel}, M. and {Possenti}, A. and {Ransom}, S. and {Theureau}, G. and {Torne}, P.},
  title         = {{A Cosmic Census of Radio Pulsars with the SKA}},
  booktitle     = {Advancing Astrophysics with the Square Kilometre Array (AASKA14)},
  year          = 2015,
  month         = apr,
  eid           = {40},
  pages         = {40},
  doi           = {10.22323/1.215.0040},
  archiveprefix = {arXiv},
  eprint        = {1501.00056},
  primaryclass  = {astro-ph.IM},
  adsurl        = {https://ui.adsabs.harvard.edu/abs/2015aska.confE..40K}
}

@article{Nan2011,
  author        = {{Nan}, Rendong and {Li}, Di and {Jin}, Chengjin and {Wang}, Qiming and {Zhu}, Lichun and {Zhu}, Wenbai and {Zhang}, Haiyan and {Yue}, Youling and {Qian}, Lei},
  title         = {{The Five-Hundred Aperture Spherical Radio Telescope (fast) Project}},
  journal       = {International Journal of Modern Physics D},
  year          = 2011,
  month         = jan,
  volume        = {20},
  number        = {6},
  pages         = {989-1024},
  doi           = {10.1142/S0218271811019335},
  archiveprefix = {arXiv},
  eprint        = {1105.3794},
  primaryclass  = {astro-ph.IM},
  adsurl        = {https://ui.adsabs.harvard.edu/abs/2011IJMPD..20..989N}
}

@article{Bower2018,
  author        = {{Bower}, Geoffrey C. and {Broderick}, Avery and {Dexter}, Jason and {Doeleman}, Shepherd and {Falcke}, Heino and {Fish}, Vincent and {Johnson}, Michael D. and {Marrone}, Daniel P. and {Moran}, James M. and {Moscibrodzka}, Monika and {Peck}, Alison and {Plambeck}, Richard L. and {Rao}, Ramprasad},
  title         = {{ALMA Polarimetry of Sgr A*: Probing the Accretion Flow from the Event Horizon to the Bondi Radius}},
  journal       = {\apj},
  year          = 2018,
  month         = {dec},
  volume        = {868},
  number        = {2},
  eid           = {101},
  pages         = {101},
  doi           = {10.3847/1538-4357/aae983},
  archiveprefix = {arXiv},
  eprint        = {1810.07317},
  primaryclass  = {astro-ph.HE},
  adsurl        = {https://ui.adsabs.harvard.edu/abs/Bower2018}
}

@article{Zhang2014,
  author        = {{Zhang}, Fupeng and {Lu}, Youjun and {Yu}, Qingjuan},
  title         = {{On the Existence of Pulsars in the Vicinity of the Massive Black Hole in the Galactic Center}},
  journal       = {\apj},
  year          = 2014,
  month         = apr,
  volume        = {784},
  number        = {2},
  eid           = {106},
  pages         = {106},
  doi           = {10.1088/0004-637X/784/2/106},
  archiveprefix = {arXiv},
  eprint        = {1402.2505},
  primaryclass  = {astro-ph.GA},
  adsurl        = {https://ui.adsabs.harvard.edu/abs/2014ApJ...784..106Z}
}

@article{DellaMonica2023d,
  author        = {{Della Monica}, Riccardo and {De Martino}, Ivan and {De Laurentis}, Mariafelicia},
  title         = {{Testing space-time geometries and theories of gravity at the Galactic centre with pulsar's time delay}},
  journal       = {\mnras},
  year          = 2023,
  month         = sep,
  volume        = {524},
  number        = {3},
  pages         = {3782-3796},
  doi           = {10.1093/mnras/stad2125},
  archiveprefix = {arXiv},
  eprint        = {2305.18178},
  primaryclass  = {gr-qc},
  adsurl        = {https://ui.adsabs.harvard.edu/abs/2023MNRAS.524.3782D}
}

@article{DeLaurentis2023,
  author        = {{De Laurentis}, Mariafelicia and {de Martino}, Ivan and {Della Monica}, Riccardo},
  title         = {{The Galactic Center as a laboratory for theories of gravity and dark matter}},
  journal       = {Reports on Progress in Physics},
  year          = 2023,
  month         = oct,
  volume        = {86},
  number        = {10},
  eid           = {104901},
  pages         = {104901},
  doi           = {10.1088/1361-6633/ace91b},
  archiveprefix = {arXiv},
  eprint        = {2211.07008},
  primaryclass  = {astro-ph.GA},
  adsurl        = {https://ui.adsabs.harvard.edu/abs/2023RPPh...86j4901D}
}

@article{Hackmann2019,
  author        = {{Hackmann}, Eva and {Dhani}, Arnab},
  title         = {{The propagation delay in the timing of a pulsar orbiting a supermassive black hole}},
  journal       = {General Relativity and Gravitation},
  year          = 2019,
  month         = mar,
  volume        = {51},
  number        = {3},
  eid           = {37},
  pages         = {37},
  doi           = {10.1007/s10714-019-2517-2},
  archiveprefix = {arXiv},
  eprint        = {1806.02547},
  primaryclass  = {gr-qc},
  adsurl        = {https://ui.adsabs.harvard.edu/abs/2019GReGr..51...37H}
}

@book{Chandrasekhar1998,
  author  = {{Chandrasekhar}, S.},
  title   = {{The Mathematical Theory of Black Holes}},
  year    = 1998,
  adsurl  = {https://ui.adsabs.harvard.edu/abs/Chandrasekhar1998}
}

@article{Damour1986,
  author   = {{Damour}, T. and {Deruelle}, N.},
  title    = {{General relativistic celestial mechanics of binary systems. II. The post-Newtonian timing formula.}},
  journal  = {Ann. Inst. Henri Poincar{\'e} Phys. Th{\'e}or},
  year     = 1986,
  month    = jan,
  volume   = {44},
  number   = {3},
  pages    = {263-292},
  adsurl   = {https://ui.adsabs.harvard.edu/abs/1986AIHS...44..263D}
}

@misc{NISTDLMF,
  key          = {DLMF, Release 1.0.26 of 2020-03-15},
  title        = {NIST Digital Library of Mathematical Functions},
  howpublished = {http://dlmf.nist.gov/, Release 1.0.26 of 2020-03-15},
  year         = {2020},
  url          = {http://dlmf.nist.gov/},
  author       = {F.~W.~J. Olver and A.~B. Olde Daalhuis and D.~W. Lozier and B.~I. Schneider and R.~F. Boisvert and C.~W. Clark and B.~R. Mille and, B.~V. Saunders and H.~S. Cohl and M.~A. McClain, eds.}
}

@article{Edwards2006,
  author        = {{Edwards}, R.~T. and {Hobbs}, G.~B. and {Manchester}, R.~N.},
  title         = {{TEMPO2, a new pulsar timing package - II. The timing model and precision estimates}},
  journal       = {\mnras},
  year          = 2006,
  month         = nov,
  volume        = {372},
  number        = {4},
  pages         = {1549-1574},
  doi           = {10.1111/j.1365-2966.2006.10870.x},
  archiveprefix = {arXiv},
  eprint        = {astro-ph/0607664},
  primaryclass  = {astro-ph},
  adsurl        = {https://ui.adsabs.harvard.edu/abs/2006MNRAS.372.1549E}
}

@article{Lai2005,
  author        = {{Lai}, Dong and {Rafikov}, Roman R.},
  title         = {{Effects of Gravitational Lensing in the Double Pulsar System J0737-3039}},
  journal       = {\apjl},
  year          = 2005,
  month         = mar,
  volume        = {621},
  number        = {1},
  pages         = {L41-L44},
  doi           = {10.1086/429146},
  archiveprefix = {arXiv},
  eprint        = {astro-ph/0411726},
  primaryclass  = {astro-ph},
  adsurl        = {https://ui.adsabs.harvard.edu/abs/2005ApJ...621L..41L}
}

@article{Taylor1979,
  author   = {{Taylor}, J.~H. and {Fowler}, L.~A. and {McCulloch}, P.~M.},
  title    = {{Measurements of general relativistic effects in the binary pulsar PSR1913 + 16}},
  journal  = {\nat},
  year     = 1979,
  month    = feb,
  volume   = {277},
  number   = {5696},
  pages    = {437-440},
  doi      = {10.1038/277437a0},
  adsurl   = {https://ui.adsabs.harvard.edu/abs/1979Natur.277..437T}
}

@article{Taylor1989,
  author   = {{Taylor}, J.~H. and {Weisberg}, J.~M.},
  title    = {{Further Experimental Tests of Relativistic Gravity Using the Binary Pulsar PSR 1913+16}},
  journal  = {\apj},
  year     = 1989,
  month    = oct,
  volume   = {345},
  pages    = {434},
  doi      = {10.1086/167917},
  adsurl   = {https://ui.adsabs.harvard.edu/abs/1989ApJ...345..434T}
}

@article{Wolszczan1992,
  author   = {{Wolszczan}, A. and {Frail}, D.~A.},
  title    = {{A planetary system around the millisecond pulsar PSR1257 + 12}},
  journal  = {\nat},
  year     = 1992,
  month    = jan,
  volume   = {355},
  number   = {6356},
  pages    = {145-147},
  doi      = {10.1038/355145a0},
  adsurl   = {https://ui.adsabs.harvard.edu/abs/1992Natur.355..145W}
}

@article{Kaspi1994,
  author   = {{Kaspi}, V.~M. and {Taylor}, J.~H. and {Ryba}, M.~F.},
  title    = {{High-Precision Timing of Millisecond Pulsars. III. Long-Term Monitoring of PSRs B1855+09 and B1937+21}},
  journal  = {\apj},
  year     = 1994,
  month    = jun,
  volume   = {428},
  pages    = {713},
  doi      = {10.1086/174280},
  adsurl   = {https://ui.adsabs.harvard.edu/abs/1994ApJ...428..713K}
}

@article{Edwards2001,
  author        = {{Edwards}, R.~T. and {van Straten}, W. and {Bailes}, M.},
  title         = {{A Search for Submillisecond Pulsars}},
  journal       = {\apj},
  year          = 2001,
  month         = oct,
  volume        = {560},
  number        = {1},
  pages         = {365-370},
  doi           = {10.1086/322772},
  archiveprefix = {arXiv},
  eprint        = {astro-ph/0106353},
  primaryclass  = {astro-ph},
  adsurl        = {https://ui.adsabs.harvard.edu/abs/2001ApJ...560..365E}
}

@article{Agazie2023,
  author        = {{Agazie}, Gabriella and {Anumarlapudi}, Akash and {Archibald}, Anne M. and {Arzoumanian}, Zaven and {Baker}, Paul T. and {B{\'e}csy}, Bence and {Blecha}, Laura and {Brazier}, Adam and {Brook}, Paul R. and {Burke-Spolaor}, Sarah and {Burnette}, Rand and {Case}, Robin and {Charisi}, Maria and {Chatterjee}, Shami and {Chatziioannou}, Katerina and {Cheeseboro}, Belinda D. and {Chen}, Siyuan and {Cohen}, Tyler and {Cordes}, James M. and {Cornish}, Neil J. and {Crawford}, Fronefield and {Cromartie}, H. Thankful and {Crowter}, Kathryn and {Cutler}, Curt J. and {Decesar}, Megan E. and {Degan}, Dallas and {Demorest}, Paul B. and {Deng}, Heling and {Dolch}, Timothy and {Drachler}, Brendan and {Ellis}, Justin A. and {Ferrara}, Elizabeth C. and {Fiore}, William and {Fonseca}, Emmanuel and {Freedman}, Gabriel E. and {Garver-Daniels}, Nate and {Gentile}, Peter A. and {Gersbach}, Kyle A. and {Glaser}, Joseph and {Good}, Deborah C. and {G{\"u}ltekin}, Kayhan and {Hazboun}, Jeffrey S. and {Hourihane}, Sophie and {Islo}, Kristina and {Jennings}, Ross J. and {Johnson}, Aaron D. and {Jones}, Megan L. and {Kaiser}, Andrew R. and {Kaplan}, David L. and {Kelley}, Luke Zoltan and {Kerr}, Matthew and {Key}, Joey S. and {Klein}, Tonia C. and {Laal}, Nima and {Lam}, Michael T. and {Lamb}, William G. and {Lazio}, T. Joseph W. and {Lewandowska}, Natalia and {Littenberg}, Tyson B. and {Liu}, Tingting and {Lommen}, Andrea and {Lorimer}, Duncan R. and {Luo}, Jing and {Lynch}, Ryan S. and {Ma}, Chung-Pei and {Madison}, Dustin R. and {Mattson}, Margaret A. and {McEwen}, Alexander and {McKee}, James W. and {McLaughlin}, Maura A. and {McMann}, Natasha and {Meyers}, Bradley W. and {Meyers}, Patrick M. and {Mingarelli}, Chiara M.~F. and {Mitridate}, Andrea and {Natarajan}, Priyamvada and {Ng}, Cherry and {Nice}, David J. and {Ocker}, Stella Koch and {Olum}, Ken D. and {Pennucci}, Timothy T. and {Perera}, Benetge B.~P. and {Petrov}, Polina and {Pol}, Nihan S. and {Radovan}, Henri A. and {Ransom}, Scott M. and {Ray}, Paul S. and {Romano}, Joseph D. and {Sardesai}, Shashwat C. and {Schmiedekamp}, Ann and {Schmiedekamp}, Carl and {Schmitz}, Kai and {Schult}, Levi and {Shapiro-Albert}, Brent J. and {Siemens}, Xavier and {Simon}, Joseph and {Siwek}, Magdalena S. and {Stairs}, Ingrid H. and {Stinebring}, Daniel R. and {Stovall}, Kevin and {Sun}, Jerry P. and {Susobhanan}, Abhimanyu and {Swiggum}, Joseph K. and {Taylor}, Jacob and {Taylor}, Stephen R. and {Turner}, Jacob E. and {Unal}, Caner and {Vallisneri}, Michele and {van Haasteren}, Rutger and {Vigeland}, Sarah J. and {Wahl}, Haley M. and {Wang}, Qiaohong and {Witt}, Caitlin A. and {Young}, Olivia and {Nanograv Collaboration}},
  title         = {{The NANOGrav 15 yr Data Set: Evidence for a Gravitational-wave Background}},
  journal       = {\apjl},
  year          = 2023,
  month         = jul,
  volume        = {951},
  number        = {1},
  eid           = {L8},
  pages         = {L8},
  doi           = {10.3847/2041-8213/acdac6},
  archiveprefix = {arXiv},
  eprint        = {2306.16213},
  primaryclass  = {astro-ph.HE},
  adsurl        = {https://ui.adsabs.harvard.edu/abs/2023ApJ...951L...8A}
}

@article{EPTA2023,
  author        = {{EPTA Collaboration} and {InPTA Collaboration} and {Antoniadis}, J. and {Arumugam}, P. and {Arumugam}, S. and {Babak}, S. and {Bagchi}, M. and {Bak Nielsen}, A. -S. and {Bassa}, C.~G. and {Bathula}, A. and {Berthereau}, A. and {Bonetti}, M. and {Bortolas}, E. and {Brook}, P.~R. and {Burgay}, M. and {Caballero}, R.~N. and {Chalumeau}, A. and {Champion}, D.~J. and {Chanlaridis}, S. and {Chen}, S. and {Cognard}, I. and {Dandapat}, S. and {Deb}, D. and {Desai}, S. and {Desvignes}, G. and {Dhanda-Batra}, N. and {Dwivedi}, C. and {Falxa}, M. and {Ferdman}, R.~D. and {Franchini}, A. and {Gair}, J.~R. and {Goncharov}, B. and {Gopakumar}, A. and {Graikou}, E. and {Grie{\ss}meier}, J. -M. and {Guillemot}, L. and {Guo}, Y.~J. and {Gupta}, Y. and {Hisano}, S. and {Hu}, H. and {Iraci}, F. and {Izquierdo-Villalba}, D. and {Jang}, J. and {Jawor}, J. and {Janssen}, G.~H. and {Jessner}, A. and {Joshi}, B.~C. and {Kareem}, F. and {Karuppusamy}, R. and {Keane}, E.~F. and {Keith}, M.~J. and {Kharbanda}, D. and {Kikunaga}, T. and {Kolhe}, N. and {Kramer}, M. and {Krishnakumar}, M.~A. and {Lackeos}, K. and {Lee}, K.~J. and {Liu}, K. and {Liu}, Y. and {Lyne}, A.~G. and {McKee}, J.~W. and {Maan}, Y. and {Main}, R.~A. and {Mickaliger}, M.~B. and {Ni{\c{t}}u}, I.~C. and {Nobleson}, K. and {Paladi}, A.~K. and {Parthasarathy}, A. and {Perera}, B.~B.~P. and {Perrodin}, D. and {Petiteau}, A. and {Porayko}, N.~K. and {Possenti}, A. and {Prabu}, T. and {Quelquejay Leclere}, H. and {Rana}, P. and {Samajdar}, A. and {Sanidas}, S.~A. and {Sesana}, A. and {Shaifullah}, G. and {Singha}, J. and {Speri}, L. and {Spiewak}, R. and {Srivastava}, A. and {Stappers}, B.~W. and {Surnis}, M. and {Susarla}, S.~C. and {Susobhanan}, A. and {Takahashi}, K. and {Tarafdar}, P. and {Theureau}, G. and {Tiburzi}, C. and {van der Wateren}, E. and {Vecchio}, A. and {Venkatraman Krishnan}, V. and {Verbiest}, J.~P.~W. and {Wang}, J. and {Wang}, L. and {Wu}, Z.},
  title         = {{The second data release from the European Pulsar Timing Array. III. Search for gravitational wave signals}},
  journal       = {\aap},
  year          = 2023,
  month         = oct,
  volume        = {678},
  eid           = {A50},
  pages         = {A50},
  doi           = {10.1051/0004-6361/202346844},
  archiveprefix = {arXiv},
  eprint        = {2306.16214},
  primaryclass  = {astro-ph.HE},
  adsurl        = {https://ui.adsabs.harvard.edu/abs/2023A&A...678A..50E}
}

@article{Reardon2023,
  author        = {{Reardon}, Daniel J. and {Zic}, Andrew and {Shannon}, Ryan M. and {Hobbs}, George B. and {Bailes}, Matthew and {Di Marco}, Valentina and {Kapur}, Agastya and {Rogers}, Axl F. and {Thrane}, Eric and {Askew}, Jacob and {Bhat}, N.~D. Ramesh and {Cameron}, Andrew and {Cury{\l}o}, Ma{\l}gorzata and {Coles}, William A. and {Dai}, Shi and {Goncharov}, Boris and {Kerr}, Matthew and {Kulkarni}, Atharva and {Levin}, Yuri and {Lower}, Marcus E. and {Manchester}, Richard N. and {Mandow}, Rami and {Miles}, Matthew T. and {Nathan}, Rowina S. and {Os{\l}owski}, Stefan and {Russell}, Christopher J. and {Spiewak}, Ren{\'e}e and {Zhang}, Songbo and {Zhu}, Xing-Jiang},
  title         = {{Search for an Isotropic Gravitational-wave Background with the Parkes Pulsar Timing Array}},
  journal       = {\apjl},
  year          = 2023,
  month         = jul,
  volume        = {951},
  number        = {1},
  eid           = {L6},
  pages         = {L6},
  doi           = {10.3847/2041-8213/acdd02},
  archiveprefix = {arXiv},
  eprint        = {2306.16215},
  primaryclass  = {astro-ph.HE},
  adsurl        = {https://ui.adsabs.harvard.edu/abs/2023ApJ...951L...6R}
}

@article{Xu2023,
  author        = {{Xu}, Heng and {Chen}, Siyuan and {Guo}, Yanjun and {Jiang}, Jinchen and {Wang}, Bojun and {Xu}, Jiangwei and {Xue}, Zihan and {Nicolas Caballero}, R. and {Yuan}, Jianping and {Xu}, Yonghua and {Wang}, Jingbo and {Hao}, Longfei and {Luo}, Jingtao and {Lee}, Kejia and {Han}, Jinlin and {Jiang}, Peng and {Shen}, Zhiqiang and {Wang}, Min and {Wang}, Na and {Xu}, Renxin and {Wu}, Xiangping and {Manchester}, Richard and {Qian}, Lei and {Guan}, Xin and {Huang}, Menglin and {Sun}, Chun and {Zhu}, Yan},
  title         = {{Searching for the Nano-Hertz Stochastic Gravitational Wave Background with the Chinese Pulsar Timing Array Data Release I}},
  journal       = {Research in Astronomy and Astrophysics},
  year          = 2023,
  month         = jul,
  volume        = {23},
  number        = {7},
  eid           = {075024},
  pages         = {075024},
  doi           = {10.1088/1674-4527/acdfa5},
  archiveprefix = {arXiv},
  eprint        = {2306.16216},
  primaryclass  = {astro-ph.HE},
  adsurl        = {https://ui.adsabs.harvard.edu/abs/2023RAA....23g5024X}
}

@article{Hobbs2006,
  author        = {{Hobbs}, G.~B. and {Edwards}, R.~T. and {Manchester}, R.~N.},
  title         = {{TEMPO2, a new pulsar-timing package - I. An overview}},
  journal       = {\mnras},
  year          = 2006,
  month         = jun,
  volume        = {369},
  number        = {2},
  pages         = {655-672},
  doi           = {10.1111/j.1365-2966.2006.10302.x},
  archiveprefix = {arXiv},
  eprint        = {astro-ph/0603381},
  primaryclass  = {astro-ph},
  adsurl        = {https://ui.adsabs.harvard.edu/abs/2006MNRAS.369..655H}
}

@article{Luo2021,
  author        = {{Luo}, Jing and {Ransom}, Scott and {Demorest}, Paul and {Ray}, Paul S. and {Archibald}, Anne and {Kerr}, Matthew and {Jennings}, Ross J. and {Bachetti}, Matteo and {van Haasteren}, Rutger and {Champagne}, Chloe A. and {Colen}, Jonathan and {Phillips}, Camryn and {Zimmerman}, Josef and {Stovall}, Kevin and {Lam}, Michael T. and {Jenet}, Fredrick A.},
  title         = {{PINT: A Modern Software Package for Pulsar Timing}},
  journal       = {\apj},
  year          = 2021,
  month         = apr,
  volume        = {911},
  number        = {1},
  eid           = {45},
  pages         = {45},
  doi           = {10.3847/1538-4357/abe62f},
  archiveprefix = {arXiv},
  eprint        = {2012.00074},
  primaryclass  = {astro-ph.IM},
  adsurl        = {https://ui.adsabs.harvard.edu/abs/2021ApJ...911...45L}
}

@article{Blandford1976,
  author  = {{Blandford}, R. and {Teukolsky}, S.~A.},
  title   = {{Arrival-time analysis for a pulsar in a binary system.}},
  journal = {\apj},
  year    = 1976,
  month   = apr,
  volume  = {205},
  pages   = {580-591},
  doi     = {10.1086/154315},
  adsurl  = {https://ui.adsabs.harvard.edu/abs/1976ApJ...205..580B}
}

@article{Kopeikin1995,
  author   = {{Kopeikin}, S.~M.},
  title    = {{On Possible Implications of Orbital Parallaxes of Wide Orbit Binary Pulsars and Their Measurability}},
  journal  = {\apjl},
  year     = 1995,
  month    = jan,
  volume   = {439},
  pages    = {L5},
  doi      = {10.1086/187731},
  adsurl   = {https://ui.adsabs.harvard.edu/abs/1995ApJ...439L...5K}
}

@article{Kopeikin1996,
  author   = {{Kopeikin}, S.~M.},
  title    = {{Proper Motion of Binary Pulsars as a Source of Secular Variations of Orbital Parameters}},
  journal  = {\apjl},
  year     = 1996,
  month    = aug,
  volume   = {467},
  pages    = {L93},
  doi      = {10.1086/310201},
  adsurl   = {https://ui.adsabs.harvard.edu/abs/1996ApJ...467L..93K}
}

@article{Wex1998,
  author        = {{Wex}, Norbert},
  title         = {{A timing formula for main-sequence star binary pulsars}},
  journal       = {\mnras},
  year          = 1998,
  month         = jul,
  volume        = {298},
  number        = {1},
  pages         = {67-77},
  doi           = {10.1046/j.1365-8711.1998.01570.x},
  archiveprefix = {arXiv},
  eprint        = {astro-ph/9706086},
  primaryclass  = {astro-ph},
  adsurl        = {https://ui.adsabs.harvard.edu/abs/1998MNRAS.298...67W}
}

@book{Lorimer2004,
  author  = {{Lorimer}, D.~R. and {Kramer}, M.},
  title   = {{Handbook of Pulsar Astronomy}},
  year    = 2004,
  volume  = {4},
  adsurl  = {https://ui.adsabs.harvard.edu/abs/2004hpa..book.....L}
}

@article{GravityCollaboration2020c,
  author        = {{Gravity Collaboration} and {Abuter}, R. and {Amorim}, A. and {Baub{\"o}ck}, M. and {Berger}, J.~P. and {Bonnet}, H. and {Brandner}, W. and {Cardoso}, V. and {Cl{\'e}net}, Y. and {de Zeeuw}, P.~T. and {Dexter}, J. and {Eckart}, A. and {Eisenhauer}, F. and {F{\"o}rster Schreiber}, N.~M. and {Garcia}, P. and {Gao}, F. and {Gendron}, E. and {Genzel}, R. and {Gillessen}, S. and {Habibi}, M. and {Haubois}, X. and {Henning}, T. and {Hippler}, S. and {Horrobin}, M. and {Jim{\'e}nez-Rosales}, A. and {Jochum}, L. and {Jocou}, L. and {Kaufer}, A. and {Kervella}, P. and {Lacour}, S. and {Lapeyr{\`e}re}, V. and {Le Bouquin}, J. -B. and {L{\'e}na}, P. and {Nowak}, M. and {Ott}, T. and {Paumard}, T. and {Perraut}, K. and {Perrin}, G. and {Pfuhl}, O. and {Rodr{\'\i}guez-Coira}, G. and {Shangguan}, J. and {Scheithauer}, S. and {Stadler}, J. and {Straub}, O. and {Straubmeier}, C. and {Sturm}, E. and {Tacconi}, L.~J. and {Vincent}, F. and {von Fellenberg}, S. and {Waisberg}, I. and {Widmann}, F. and {Wieprecht}, E. and {Wiezorrek}, E. and {Woillez}, J. and {Yazici}, S. and {Zins}, G.},
  title         = {{Detection of the Schwarzschild precession in the orbit of the star S2 near the Galactic centre massive black hole}},
  journal       = {\aap},
  year          = 2020,
  month         = {apr},
  volume        = {636},
  eid           = {L5},
  pages         = {L5},
  doi           = {10.1051/0004-6361/202037813},
  archiveprefix = {arXiv},
  eprint        = {2004.07187},
  primaryclass  = {astro-ph.GA},
  adsurl        = {https://ui.adsabs.harvard.edu/abs/GravityCollaboration2020}
}

@article{ForemanMackey2013,
  author        = {{Foreman-Mackey}, Daniel and {Hogg}, David W. and {Lang}, Dustin and {Goodman}, Jonathan},
  title         = {{emcee: The MCMC Hammer}},
  journal       = {\pasp},
  year          = 2013,
  month         = mar,
  volume        = {125},
  number        = {925},
  pages         = {306},
  doi           = {10.1086/670067},
  archiveprefix = {arXiv},
  eprint        = {1202.3665},
  primaryclass  = {astro-ph.IM},
  adsurl        = {https://ui.adsabs.harvard.edu/abs/2013PASP..125..306F}
}

@incollection{Figer2009,
  author    = {{Figer}, Don. F.},
  title     = {{Massive-star formation in the Galactic center:}},
  booktitle = {Massive Stars: From Pop III and GRBs to the Milky Way. Space Telescope Science Institute Symposium Series No. 20. Edited by Mario Livio and Eva Villaver. Cambridge University Press},
  year      = 2009,
  editor    = {{Livio}, Mario and {Villaver}, Eva},
  pages     = {40-59},
  doi       = {10.1017/CBO9780511770593.004},
  adsurl    = {https://ui.adsabs.harvard.edu/abs/2009msfp.book...40F}
}

@ARTICLE{GravityCollaboration2024,
       author = {{Gravity Collaboration} and {Abd El Dayem}, K. and {Abuter}, R. and {Aimar}, N. and {Amaro Seoane}, P. and {Amorim}, A. and {Beck}, J. and {Berger}, J.~P. and {Bonnet}, H. and {Bourdarot}, G. and {Brandner}, W. and {Cardoso}, V. and {Capuzzo Dolcetta}, R. and {Cl{\'e}net}, Y. and {Davies}, R. and {de Zeeuw}, P.~T. and {Drescher}, A. and {Eckart}, A. and {Eisenhauer}, F. and {Feuchtgruber}, H. and {Finger}, G. and {F{\"o}rster Schreiber}, N.~M. and {Foschi}, A. and {Gao}, F. and {Garcia}, P. and {Gendron}, E. and {Genzel}, R. and {Gillessen}, S. and {Hartl}, M. and {Haubois}, X. and {Haussmann}, F. and {Hei{\ss}el}, G. and {Henning}, T. and {Hippler}, S. and {Horrobin}, M. and {Jochum}, L. and {Jocou}, L. and {Kaufer}, A. and {Kervella}, P. and {Lacour}, S. and {Lapeyr{\`e}re}, V. and {Le Bouquin}, J. -B. and {L{\'e}na}, P. and {Lutz}, D. and {Mang}, F. and {More}, N. and {Ott}, T. and {Paumard}, T. and {Perraut}, K. and {Perrin}, G. and {Pfuhl}, O. and {Rabien}, S. and {Ribeiro}, D.~C. and {Sadun Bordoni}, M. and {Scheithauer}, S. and {Shangguan}, J. and {Shimizu}, T. and {Stadler}, J. and {Straub}, O. and {Straubmeier}, C. and {Sturm}, E. and {Tacconi}, L.~J. and {Urso}, I. and {Vincent}, F. and {von Fellenberg}, S.~D. and {Widmann}, F. and {Wieprecht}, E. and {Woillez}, J. and {Zhang}, F.},
        title = "{Improving constraints on the extended mass distribution in the Galactic center with stellar orbits}",
      journal = {\aap},
         year = 2024,
        month = dec,
       volume = {692},
          eid = {A242},
        pages = {A242},
          doi = {10.1051/0004-6361/202452274},
archivePrefix = {arXiv},
       eprint = {2409.12261},
 primaryClass = {astro-ph.GA},
       adsurl = {https://ui.adsabs.harvard.edu/abs/2024A&A...692A.242G}
}

@ARTICLE{PyGRO2025,
       author = {{Della Monica}, Riccardo},
        title = "{PyGRO: A Python integrator for General Relativistic Orbits}",
      journal = {\aap},
         year = 2025,
        month = jun,
       volume = {698},
          eid = {A193},
        pages = {A193},
          doi = {10.1051/0004-6361/202554300},
archivePrefix = {arXiv},
       eprint = {2504.20152},
 primaryClass = {gr-qc},
       adsurl = {https://ui.adsabs.harvard.edu/abs/2025A&A...698A.193D}
}

@ARTICLE{Rees1988,
       author = {{Rees}, Martin J.},
        title = "{Tidal disruption of stars by black holes of {}10$^{6}$-{}10$^{8}$ solar masses in nearby galaxies}",
      journal = {\nat},
         year = 1988,
        month = jun,
       volume = {333},
       number = {6173},
        pages = {523-528},
          doi = {10.1038/333523a0},
       adsurl = {https://ui.adsabs.harvard.edu/abs/1988Natur.333..523R}
}

@ARTICLE{Generozov2025,
       author = {{Generozov}, Aleksey and {Perets}, Hagai B. and {Bordoni}, Matteo S. and {Bourdarot}, Guillaume and {Drescher}, Antonia and {Eisenhauer}, Frank and {Genzel}, Reinhard and {Gillessen}, Stefan and {Mang}, Felix and {Ott}, Thomas and {Ribeiro}, Diogo C. and {Sch{\"o}del}, Rainer},
        title = "{The S stars' zone of avoidance in the Galactic center}",
      journal = {\aap},
         year = 2025,
        month = apr,
       volume = {696},
          eid = {A68},
        pages = {A68},
          doi = {10.1051/0004-6361/202453272},
archivePrefix = {arXiv},
       eprint = {2412.02752},
 primaryClass = {astro-ph.GA},
       adsurl = {https://ui.adsabs.harvard.edu/abs/2025A&A...696A..68G}
}

@ARTICLE{Waisberg2018,
       author = {{Waisberg}, Idel and {Dexter}, Jason and {Gillessen}, Stefan and {Pfuhl}, Oliver and {Eisenhauer}, Frank and {Plewa}, Phillip M. and {Baub{\"o}ck}, Michi and {Jimenez-Rosales}, Alejandra and {Habibi}, Maryam and {Ott}, Thomas and {von Fellenberg}, Sebastiano and {Gao}, Feng and {Widmann}, Felix and {Genzel}, Reinhard},
        title = "{What stellar orbit is needed to measure the spin of the Galactic centre black hole from astrometric data?}",
      journal = {\mnras},
         year = 2018,
        month = may,
       volume = {476},
       number = {3},
        pages = {3600-3610},
          doi = {10.1093/mnras/sty476},
archivePrefix = {arXiv},
       eprint = {1802.08198},
 primaryClass = {astro-ph.GA},
       adsurl = {https://ui.adsabs.harvard.edu/abs/2018MNRAS.476.3600W}
}

@ARTICLE{Chen2023,
       author = {{Chen}, Zhuo and {Do}, Tuan and {Ghez}, Andrea M. and {Hosek}, Matthew W. and {Feldmeier-Krause}, Anja and {Chu}, Devin S. and {Bentley}, Rory O. and {Lu}, Jessica R. and {Morris}, Mark R.},
        title = "{The Star Formation History of the Milky Way's Nuclear Star Cluster}",
      journal = {\apj},
         year = 2023,
        month = feb,
       volume = {944},
       number = {1},
          eid = {79},
        pages = {79},
          doi = {10.3847/1538-4357/aca8ad},
archivePrefix = {arXiv},
       eprint = {2212.01397},
 primaryClass = {astro-ph.GA},
       adsurl = {https://ui.adsabs.harvard.edu/abs/2023ApJ...944...79C}
}

@ARTICLE{Shannon2017,
       author = {{Shannon}, R.~M. and {Cordes}, J.~M.},
        title = "{Modelling and mitigating refractive propagation effects in precision pulsar timing observations}",
      journal = {\mnras},
         year = 2017,
        month = jan,
       volume = {464},
       number = {2},
        pages = {2075-2089},
          doi = {10.1093/mnras/stw2449},
archivePrefix = {arXiv},
       eprint = {1609.07573},
 primaryClass = {astro-ph.IM},
       adsurl = {https://ui.adsabs.harvard.edu/abs/2017MNRAS.464.2075S}
}

@ARTICLE{Rickett1977,
       author = {{Rickett}, B.~J.},
        title = "{Interstellar scattering and scintillation of radio waves.}",
      journal = {\araa},
         year = 1977,
        month = jan,
       volume = {15},
        pages = {479-504},
          doi = {10.1146/annurev.aa.15.090177.002403},
       adsurl = {https://ui.adsabs.harvard.edu/abs/1977ARA&A..15..479R}
}

@ARTICLE{Kramer2021,
       author = {{Kramer}, M. and {Stairs}, I.~H. and {Manchester}, R.~N. and {Wex}, N. and {Deller}, A.~T. and {Coles}, W.~A. and {Ali}, M. and {Burgay}, M. and {Camilo}, F. and {Cognard}, I. and {Damour}, T. and {Desvignes}, G. and {Ferdman}, R.~D. and {Freire}, P.~C.~C. and {Grondin}, S. and {Guillemot}, L. and {Hobbs}, G.~B. and {Janssen}, G. and {Karuppusamy}, R. and {Lorimer}, D.~R. and {Lyne}, A.~G. and {McKee}, J.~W. and {McLaughlin}, M. and {M{\"u}nch}, L.~E. and {Perera}, B.~B.~P. and {Pol}, N. and {Possenti}, A. and {Sarkissian}, J. and {Stappers}, B.~W. and {Theureau}, G.},
        title = "{Strong-Field Gravity Tests with the Double Pulsar}",
      journal = {Physical Review X},
         year = 2021,
        month = oct,
       volume = {11},
       number = {4},
          eid = {041050},
        pages = {041050},
          doi = {10.1103/PhysRevX.11.041050},
archivePrefix = {arXiv},
       eprint = {2112.06795},
 primaryClass = {astro-ph.HE},
       adsurl = {https://ui.adsabs.harvard.edu/abs/2021PhRvX..11d1050K}
}

@ARTICLE{Hu2022,
       author = {{Hu}, H. and {Kramer}, M. and {Champion}, D.~J. and {Wex}, N. and {Parthasarathy}, A. and {Pennucci}, T.~T. and {Porayko}, N.~K. and {van Straten}, W. and {Venkatraman Krishnan}, V. and {Burgay}, M. and {Freire}, P.~C.~C. and {Manchester}, R.~N. and {Possenti}, A. and {Stairs}, I.~H. and {Bailes}, M. and {Buchner}, S. and {Cameron}, A.~D. and {Camilo}, F. and {Serylak}, M.},
        title = "{Gravitational signal propagation in the double pulsar studied with the MeerKAT telescope}",
      journal = {\aap},
         year = 2022,
        month = nov,
       volume = {667},
          eid = {A149},
        pages = {A149},
          doi = {10.1051/0004-6361/202244825},
archivePrefix = {arXiv},
       eprint = {2209.11798},
 primaryClass = {astro-ph.HE},
       adsurl = {https://ui.adsabs.harvard.edu/abs/2022A&A...667A.149H}
}

@ARTICLE{Damour1988,
       author = {{Damour}, T. and {Schafer}, G.},
        title = "{Higher-order relativistic periastron advances and binary pulsars.}",
      journal = {Nuovo Cimento B Serie},
         year = 1988,
        month = jan,
       volume = {101B},
       number = {2},
        pages = {127-176},
          doi = {10.1007/BF02828697},
       adsurl = {https://ui.adsabs.harvard.edu/abs/1988NCimB.101..127D}
}

@ARTICLE{Manchester2005,
       author = {{Manchester}, R.~N. and {Hobbs}, G.~B. and {Teoh}, A. and {Hobbs}, M.},
        title = "{The Australia Telescope National Facility Pulsar Catalogue}",
      journal = {\aj},
         year = 2005,
        month = apr,
       volume = {129},
       number = {4},
        pages = {1993-2006},
          doi = {10.1086/428488},
archivePrefix = {arXiv},
       eprint = {astro-ph/0412641},
 primaryClass = {astro-ph},
       adsurl = {https://ui.adsabs.harvard.edu/abs/2005AJ....129.1993M}
}

\appendix

\section{Convergence and validation of the interpolation pipeline}
\label{app:interpolation}

\begin{figure}[t]
  \includegraphics[width=\columnwidth]{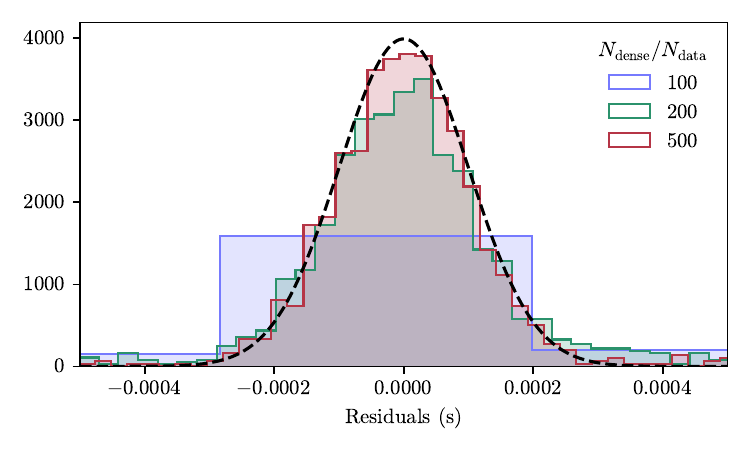}
  \caption{Residuals (in seconds) obtained by comparing the observed TOAs (reference timing solution + Gaussian noise $\sigma_{\rm TOA}=100\,\mu$s) with the model prediction derived from the quintic-spline interpolation for different oversampling factors $N_{\rm dense}/N_{\rm data}$. When $N_{\rm dense}=500\,N_{\rm data}$ the residual distribution becomes Gaussian with variance matching the instrumental noise, indicating that interpolation errors are negligible. This panel illustrates the results for Toy Model 2; the other toy models show consistent behaviour.} \label{fig:app_hist}
\end{figure}

\begin{figure}[t]
  \includegraphics[width=\columnwidth]{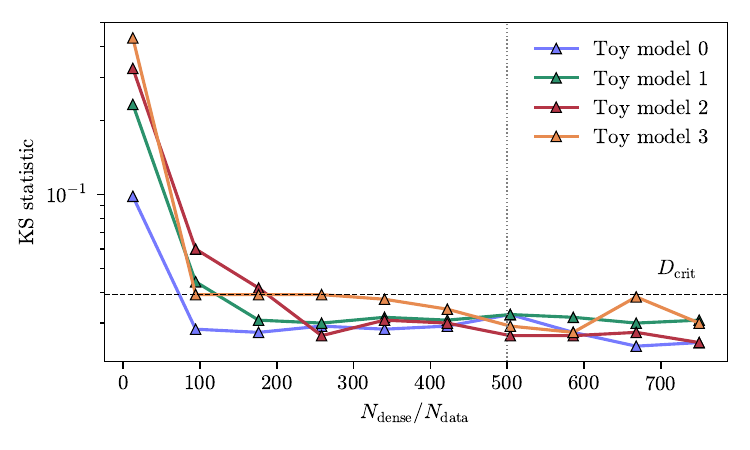}
  \caption{For each toy model we compute the KS statistic $D$ comparing the residuals distribution to a Gaussian with $\sigma_{\rm TOA}=100\,\mu$s. The horizontal line marks the critical value $D_{\rm crit}$ at the $5\%$ significance level (computed as $1.36/\sqrt{N_{\rm data}}$). Convergence is reached when $D\le D_{\rm crit}$; for all toy models considered the choice $N_{\rm dense}=500\,N_{\rm data}$ is sufficient.} \label{fig:app_ks}
\end{figure}

Here we detail the numerical choices, error scalings and the convergence test used to validate the interpolation of the timing function $\textrm{TOA}(\tau)$ employed in the manuscript. The goals are \emph{(i)} to demonstrate that the interpolation does not introduce phase-dependent systematics in the timing residuals, and \emph{(ii)} to justify the adopted oversampling factor $N_{\rm dense}=500\,N_{\rm data}$.

The time-like geodesic equations are integrated using a Dormand–Prince 5(4) solver (DOPRI5), implemented in \texttt{PyGRO} \cite{PyGRO2025}. Since the algorithm uses local extrapolation, the per-step \emph{local} truncation error of an order-$p$ explicit RK method scales as $\mathcal{O}(h^{p+1})$. For DOPRI5 this gives a per-step local error $\mathcal{O}(h^{6})$. On the other hand, the \emph{global} truncation error accumulated over an interval scales as $\mathcal{O}(h^{p})$. For DOPRI5 the global error therefore scales as $\mathcal{O}(h^{5})$. Our implementation exploits the integrator's dense output capability: at each step the integrator can provide a continuous polynomial representation of the solution inside the step, allowing us to evaluate the emitter position and velocity at arbitrary proper times with the same order-of-accuracy guaranteed by the integrator and the chosen integration tolerance. Because of the dense output, we do not interpolate the geometric trajectory of the emitter; the emitter's coordinates and four-velocity are available at the integration tolerance level for any queried proper time. The only interpolated object in our pipeline is the forward timing function $\textrm{TOA}(\tau)$ representing the coordinate time of arrival as a function of emitter proper time. To interpolate $\textrm{TOA}(\tau)$ we construct a quintic spline (polynomial degree $d=5$) from sampled values of $\textrm{TOA}(\tau)$ obtained via dense-output evaluations of the integrator. Let us recall that for a spline of degree $d$ constructed on a uniform grid with spacing $h$, and for sufficiently smooth $f(\tau)$ (i.e., $f^{(d+1)}$ bounded), classical approximation results yield an interpolation error bound of the form
\begin{equation}
    \| f - S_d\|_\infty \le C \, h^{d+1} \, \max_{\xi\in[\tau_{\min},\tau_{\max}]} |f^{(d+1)}(\xi)|,
\end{equation}
being $S_d$ the constructed spline, for some constant $C$ depending on the spline construction but not on $h$. For the quintic spline this implies that the interpolation errors scale as $\mathcal{O}(h^{6})$. Therefore, for a given sampling spacing  $h$, the interpolation error of $\textrm{TOA}(\tau)$ is at least as small as the per-step local truncation error and is \emph{generically smaller} than the global integration error; consequently interpolation is not expected to be the leading source of global error provided $\textrm{TOA}(\tau)$ is sufficiently smooth at the chosen resolution. 

To demonstrate that the interpolation strategy does not inject phase-dependent systematics, we adopt an empirical protocol based on the distribution of timing residuals. Since we are injecting gaussian noise with a known amplitude in our mock catalogues, we know a priori that the residuals distribution in case of a perfect sampling of the inverse timing formula is $\mathcal{N}(0,\sigma_{\rm TOA}^2)$, namely a gaussian with zero mean and standard deviation $\sigma_{\rm TOA}=100\,\mu$s for SKA-like sensitivity. In other words, if the inverse timing formula is correctly interpolated, the residuals are noise-dominated. Conversely, if we introduce spurious effects in the inversion of the timing function, these will appear in the residuals which will not be anymore noise-dominate and their distribution will thus deviate from $\mathcal{N}(0,\sigma_{\rm TOA}^2)$. Hence, for our convergence check, for a chosen oversampling factor $N_{\rm dense}$ we compare the empirical cumulative distribution $F_N$ of the residuals to the cumulative distribution $F_0$ of the reference Gaussian $\mathcal{N}(0,\sigma_{\rm TOA}^{2})$ using the one-sample Kolmogorov–Smirnov (KS) statistic
\begin{equation}
    D = \sup_x \big|F_N(x) - F_0(x)\big|.
\end{equation}
For a sample of size $N_{\rm data}$, the critical value at a given significance level $\alpha$ is
\begin{equation}
    D_{\rm crit}(\alpha) = \frac{c(\alpha)}{\sqrt{N_{\rm data}}},
\end{equation}
with $c(0.05)\approx 1.36$ for $\alpha=0.05$. If $D \le D_{\rm crit}(0.05)$, we accept the null hypothesis that the residuals are drawn from the reference Gaussian at the 5\% level, i.e. the residuals are statistically indistinguishable from instrument noise alone.

Using the protocol above we vary $N_{\rm dense}$ and compute $D(N_{\rm dense})$ for all the toy models introduced in Table \ref{tab:pulsars_toy_models}. The two representative diagnostics are reported in Fig.~\ref{fig:app_hist} (histogram of residuals for different $N_{\rm dense}$, for the specific case of the Toy model 2) and Fig.~\ref{fig:app_ks} (KS statistic as a function of $N_{\rm dense}$). The figures demonstrate that \emph{(i)} for small $N_{\rm dense}$ interpolation/inversion errors are visible as non-Gaussian tails or as broadened distributions in the residuals histogram; the KS statistic $D$ exceeds the critical threshold $D_{\rm crit}$, rejecting the hypothesis of Gaussian, noise-dominated residuals; $(ii)$ as $N_{\rm dense}$ increases, the distribution of residuals becomes indistinguishable from the reference Gaussian and the KS statistic falls below $D_{\rm crit}$. For all toy models considered in the manuscript the choice $N_\textrm{dense}=500\,N_\textrm{data}$ yields residuals that pass the KS test and are therefore consistent with being dominated by instrument noise (no interpolation-induced phase structure). We thus adopt this oversampling factor in the simulations presented throughout this work. Empirically, the interpolation/inversion errors for the adopted $N_{\rm dense}$ are at least an order of magnitude smaller than the SKA goal $\sigma_{\rm TOA}=100\,\mu$s; therefore interpolation does not limit the timing accuracy in our experiments.

\end{document}